\documentclass[aps,prd,eqsecnum,12pt,showpacs,preprintnumbers,nofootinbib,superscriptaddress]{revtex4-2}
\usepackage{geometry}                
\usepackage[pdftex]{graphicx}
\usepackage{float}
\usepackage{amssymb,amsmath,amsopn,bm,dsfont,mathrsfs}

\usepackage[scr]{rsfso}
\DeclareMathAlphabet{\mathpzc}{OT1}{pzc}{m}{it}

\newcommand{\sI}{\mathscr{I}}

\usepackage[colorlinks,linkcolor=blue, pdfstartview=FitH]{hyperref}
\usepackage{enumitem}
\usepackage{txfonts}
\usepackage{wrapfig}
\usepackage[compact,small]{titlesec}
\usepackage{amsbsy}

\hypersetup{linkcolor=blue,citecolor=blue,filecolor=black,urlcolor=blue}

\IfFileExists{dsfont.sty}
	{\usepackage{dsfont}
         \let\mathbb=\mathds
         \newcommand{\id}{\mathds{1}}}
	{\typeout{Package dsfont.sty was not found, using alternative macros.}
         \let\mathds=\mathbb
         \newcommand{\id}{\mbox{1 \kern-.59em \textrm{l}}}}

\renewcommand\a{\alpha}
\renewcommand\b{\beta}
\renewcommand\d{\delta}
\newcommand\ka{\kappa}

\renewcommand\r{\rho}

\renewcommand\t{\tau}

\renewcommand\j{\psi}
\renewcommand\th{\theta}

\newcommand\e{\epsilon}
\newcommand\g{\gamma}

\newcommand\m{\mu}
\newcommand\n{\nu}

\newcommand\p{\pi}
\newcommand\h{\eta}
\newcommand\s{\sigma}
\newcommand\f{\phi}
\newcommand\w{\omega}

\newcommand\vf{\varphi}

\newcommand\tu{\tilde u}
\newcommand\tv{\tilde \vv}

\newcommand{\vv}{\varv}

\newcommand{\ha}{\hat a}
\newcommand{\hb}{\hat b}
\newcommand{\hc}{\hat c}
\newcommand{\hd}{\hat d}

\newcommand{\hu}{\hat u}
\newcommand{\hv}{\hat \varv}
\newcommand{\hth}{\hat \theta}
\newcommand{\hf}{\hat \phi}

\renewcommand\P{\Pi}

\newcommand\D{\Delta}

\newcommand\W{\Omega}

\newcommand{\lag}{\langle}
\newcommand{\rag}{\rangle}

\newcommand{\cA}{{\cal A}}
\newcommand{\cC}{{\cal C}}
\newcommand{\cD} {{\mathcal D}}

\newcommand{\cO}{{\cal O}}

\newcommand{\cR}{{\cal R}}

\newcommand{\cT}{{\cal T}}

\newcommand{\pa}{\partial}

\newcommand{\nn}{\nonumber \\}
\newcommand{\na}{\nabla}
\newcommand{\sdfrac}[2]{\mbox{\small$\displaystyle\frac{#1}{#2}$}}

\newcommand{\rM}{r_{\!_M}}
\newcommand{\bea}{\begin{eqnarray}}
\newcommand{\eea}{\end{eqnarray}}
\newcommand{\be}{\begin{equation}}
\newcommand{\ee}{\end{equation}}
\newcommand{\bes}{\begin{subequations}}
\newcommand{\ees}{\end{subequations}}

\def\nbox#1#2{\vcenter{\hrule \hbox{\vrule height#2in
\kern#1in \vrule} \hrule}}
\def\sq{\,\raise0.8pt\hbox{$\nbox{.10}{.10}$}\,}
\def\sqb{\,\raise.5pt\hbox{$\overline{\nbox{.09}{.09}}$}\,}

\usepackage[T1]{fontenc} 
\usepackage[titletoc]{appendix}
\usepackage{setspace}

\allowdisplaybreaks

\begin{document}
\pagestyle{empty}

\title{Quantum Effects of the Conformal Anomaly\\[-1ex] in a 2D Model of Gravitational Collapse}

\author{Emil Mottola}
\email{mottola.emil@gmail.com, emottola@unm.edu}
\affiliation{Dept.~of Physics and Astronomy, Univ.~of New Mexico,
Albuquerque NM 87131 USA}

\author{Mani Chandra}
\email{mc0710@gmail.com}
\affiliation{Dept.~of Materials Science and Engineering,\\[-1.5ex]
Rensselaer Polytechnic Institute, Troy, NY 12180 USA}

\author{Gian Mario Manca}
\email{gianmariomanca@gmail.com}
\affiliation{Max Planck Institute for Gravitationsphysik, Albert Einstein Institute,\\[-1.5ex]
Am Mühlenberg 1, 14476 Potsdam, Germany and\\[-1.5ex]
Callinstra\ss e 38, 30167 Hannover, Germany}

\author{Evgeny Sorkin}
\email{evg.sorkin@gmail.com }
\affiliation{Perceive Corp., San Jose, CA 95002, USA\\[-1.5ex]
and Vancouver, BC V6C 1N5, Canada}

\begin{abstract}
\vspace{1mm}
The macroscopic effects of the quantum conformal anomaly are evaluated in a simplified two-dimensional model 
of gravitational collapse. The effective action and stress tensor of the anomaly can be expressed in a local quadratic 
form by the introduction of a scalar conformalon field $\vf$, which satisfies a linear wave equation. A wide class of 
non-vacuum initial state conditions is generated by different solutions of this equation. An interesting subclass of solutions 
corresponds to initial states that give rise to an arbitrarily large semi-classical stress tensor $\big\lag T\!_\m^{\ \,\n}\big\rag$ 
on the future horizon of the black hole formed in classical collapse. These lead to modification and suppression of Hawking 
radiation at late times after the collapse, and potentially large backreaction effects on the horizon scale due to the
conformal anomaly. The probability of non-vacuum initial conditions large enough to produce these effects is 
estimated from the Gaussian vacuum wave functional of $\vf$ in the Schr\"odinger representation and shown to be 
$\cO(1)$. These results indicate that quantum effects of the conformal anomaly in non-vacuum states are relevant 
for gravitational collapse in the effective theory of gravity in four dimensions as well.
\end{abstract} 

\maketitle
\vfil
\eject

\tableofcontents
\pagenumbering{arabic}
\pagestyle{plain}
\vfil\break

\section{Introduction}
\label{Sec:Intro}

Black holes are solutions of the Einstein eqs.~of classical general relativity (GR) in the absence of sources, except
for interior singularities where matter is compressed to infinite pressures and densities. In addition to these
singularities, the characteristic feature of a classical black hole (BH) is its event horizon, 
the critical null surface of finite area from which outwardly directed light rays cannot escape. 

Whereas it is widely believed that quantum effects intervene to regulate interior BH singularities, the horizon 
region is generally supposed to remain substantially unchanged from the classical description. This description 
includes the important, but often unstated assumption, of vanishing stress tensor $T_{\!\m\n}\!=\!0$ on the horizon
that permits continuation of the exterior geometry into the BH interior by means of a (singular)
transformation of coordinates~\cite{MTW,HawkEllis:1973}. 

It is important to critically examine this assumption for a number of reasons. Even in classical GR, the hyperbolic 
character of Einstein's eqs.~allows generically for $T_{\!\m\n}$ sources and discontinuities on the horizon which 
would violate the hypothesis of analytic continuation through it, potentially altering the geometry of the singular 
interior as well. Critical examination of assumptions about the stress tensor on the horizon is all the more warranted 
when quantum effects are considered. If the quantum state is assumed to be the local vacuum at the horizon, the 
expectation value of the stress tensor $\lag T\!_\m^{\ \,\n}\rag$ in this state can remain negligibly small, but only 
provided that quantum fluctuations measured by higher point correlation functions such as $\lag T_{\!\a\b} T_{\!\m\n}\rag$ 
also remain small on the horizon. This condition in particular is very much open to question in the quantum theory,
as we shall discuss in this paper.

Regarding the quantum state on the horizon, it is well known that there is no unique vacuum state in curved spacetime~\cite{BirDav}. 
In flat Minkowski space the existence of a unique vacuum ground state relies upon the Lorentz invariant separation of positive 
and negative frequency modes, hence particle and anti-particle states, over a complete Cauchy surface, and the existence of a positive 
definite Hamiltonian with respect to that hypersurface. These requirements are not satisfied in general curved spacetimes, and are 
particularly problematic when horizons are present. At a BH horizon the timelike Killing field $\pa_t$ (or the co-rotating Killing field 
$\pa_t+ \w\,\pa_\f$ for rotating BHs) becomes null, and the clean separation of particle and anti-particle modes breaks down, while 
beyond the horizon the Killing norm changes sign and the corresponding Hamiltonian becomes unbounded from below. There is thus 
no {\it a priori} reason for the state of QFT to correspond to the `empty' Minkowski vacuum at the horizon, or for quantum 
fluctuations from that state to remain small there. Certainly a large variety of non-vacuum states with $\lag T\!_\m^{\ \,\n}\rag\!\neq\! 0$ 
are also allowed, and can be considered.

Early work established that the Hawking effect is dependent upon this choice of quantum state, and is also closely related 
to the conformal anomaly that arises in defining the renormalized $\lag T_{\m\n}\rag$ in BH  
spacetimes~\cite{DavFulUnrPRD76,ChrFul:1977}. Later it was shown that Hawking thermal emission at late times 
after gravitational collapse to a BH can be derived directly from the assumption that the short distance properties 
of the quantum state and the Hadamard behavior of its Green's functions on the future horizon region are the same 
as those in flat space~\cite{FredHag:1990}. This assumption also guarantees that the future horizon is smooth, and 
$\lag T\!_{\m}^{\ \,\n}\rag$ remains regular there, so that quantum backreaction effects remain small. These conditions 
correspond to the initial state of QFT in gravitational collapse to be the Unruh state~\cite{Unruh:1976}. Virtually all later
investigations have assumed this state, including those with dynamical backreaction~\cite{CGHSPRD92,ParPirPRL94}.

It is also the regularity of the horizon and absence of any stress tensor source there that allows association
of a temperature $T_{\!_H} = 1/\b_{_H}$ with the periodicity $\b_{_H}$ of the metric at the horizon continued 
to Euclidean time~\cite{HartHawk:1976,GibPerRSL78}. Yet paradoxically, it is just this assumption 
of a smooth horizon and the Hawking temperature associated to it that leads to an enormous Bekenstein-Hawking 
BH entropy equal to $1/4$ of the area of the horizon, which is particularly difficult to understand if the BH horizon 
is a smooth mathematical boundary only, with no sources or independent degrees of freedom of its own. If matter 
and information can freely fall just one-way through this mathematical horizon boundary, the effect of Hawking 
thermal radiation also suggests the possibility of pure states evolving into mixed states and the breakdown of quantum 
unitary evolution~\cite{HawkUnit:1976}. The difficulty, if not impossibility, of recovering this lost information at the 
late or final stages of the BH evaporation process leads to a severe `information paradox,' that has been the subject 
of numerous investigations and speculations spanning several decades~\cite{Preskill:1992,
Page:1993,Mathur:2009,AMPS:2013,Giddings:2013,MottVauPT,Marolf:2017,UnruhWald:2017,AHMST:RMP21}. 

Although the Hawking temperature $T_{\!_H}$ of radiation far from the BH is very small, the inverse of the gravitational 
redshift implies infinitely blueshifted local temperatures and energies if traced back to the horizon. It is thus by no means 
clear that quantum fluctuations $\lag T_{\!\a\b} T_{\!\m\n}\rag$ from the mean and their backreaction on the near-horizon 
geometry can be neglected, as is usually assumed. The increasing time dilation and gravitational blueshift of frequency 
and energy scales with respect to the asymptotically flat region as the horizon is approached results in all fixed finite 
mass scales becoming negligible there, and an effective classical conformal symmetry in the near-horizon 
region~\cite{SacSol:2001,MazEMWeyl:2001,AntMazEM:2012}. This implies that the conformal 
behavior and conformal anomaly of QFT are relevant there~\cite{EMVau:2006,EMZak:2010,EMEFT:2022}. 

It is also known that the conformal anomaly is necessarily associated with the existence and residue of a $1/k^2$ 
massless pole in stress tensor correlation functions, even in flat space~\cite{BertKohl:AP01,GiaEM:2009,EMZak:2010,BlaCabEM:2014,
TTTCFT:2019}. Since this massless anomaly pole in quantum correlation functions is a lightlike singularity, it
is associated with effects on the light cone, which can extend to arbitrarily large macroscopic scales, and is particularly 
relevant on null horizons. The $1/k^2$ pole can be expressed as the propagator of an effective scalar degree of freedom $\vf$, 
a collective {\it conformalon} mode of the underlying massless (or sufficiently light) quantum fields, whose fluctuations 
and correlations are significantly enhanced in the vicinity of a BH horizon. The existence of a lightlike singularity implies 
quantum correlations due to the anomaly which influence the semi-classical mean value $\lag T_{\!\m}^{\ \n}\rag$
as well. The dependence of the long range conformalon scalar on the norm of the Killing vector $\pa_t$ carries non-local 
information about the conformal transformation of the vacuum from the asymptotically flat region where the Minkowski 
vacuum is preferred, to the expectation value $\lag T_{\!\m}^{\ \n}\rag$ on the BH horizon. 

These quantum anomaly effects on the horizon are generically large for wide classes of non-vacuum initial conditions, 
notwithstading the smallness of the curvature there~\cite{EMVau:2006,AndEMVau:PRD07,EMZak:2010}. The local form 
of the anomaly effective action and stress tensor in terms of the scalar $\vf$ makes the quantitative evaluation of these 
effects much simpler technically than the much more involved and laborious method of obtaining renormalized expectations 
values $\lag T_{\!\m}^{\ \n}\rag$ directly from the underlying QFT~\cite{AndHisSam:1995}. Indeed the technical 
complexity of the direct method of calculating $\lag T_{\!\m}^{\ \n}\rag$ has been sufficient to deter any systematic 
investigation of all but a small number of special quantum states, in specific QFTs. 

In contrast, a very wide class of states in generic conformal QFTs can be investigated by simply considering the variety 
of possible solutions to the {\it linear} wave eq.~satisfied by the conformalon scalar $\vf$ field, and computing its 
semi-classical $T_{\!\m}^{\ \n}[\vf]$, which is already renormalized. Since the corresponding effective action of the 
anomaly is also quadratic in $\vf$, any particular occurrence of non-vacuum initial data in gravitational collapse is described 
by a Gaussian wavefunctional in the Schr\"odinger representation, and its probability is therefore also easily estimated. 
Because all of these essential features are present in both two and four spacetime dimensions, it is advantageous 
to investigate their consequences first in the 2D case, in a simplified computable model of gravitational collapse 
without backreaction, as a proxy and warm-up to the more realistic 4D problem. 

With this purpose in mind, the organization of the paper is as follows. In the next section we define the two-dimensional model, 
and set notations and conventions in double null coordinates suitable for gravitational collapse. In Sec.~\ref{Sec:ClCollapse} we
specify and solve for the interior and exterior geometry of an imploding null shell which creates a classical BH.
In Sec.~\ref{Sec:2DAnom} we review the two-dimensional conformal anomaly and non-local Polyakov effective
action corresponding to it, the massless pole it generates in vacuum polarization, and the local representation 
of the effective action by the introduction of the massless scalar conformalon field $\vf$, showing
how it can have significant effects on BH horizons. In Sec.~\ref{Sec:Init} we evaluate the anomaly stress 
tensor $T_{\!\m}^{\ \n}[\vf]$ in a subclass of interesting non-vacuum states where it can become arbitrarily large
and suppress the Hawking effect. In Sec.~\ref{Sec:Prob} we make use of the Gaussian distribution corresponding 
to these initial states in the wavefunctional of the anomaly effective action to show that the probability of non-vacuum 
initial conditions producing such effects on the horizon are non-negligible and $\cO(1)$, showing also
how this is consistent with general theorems of finite initial data, such as~\cite{FulSweWald:CMP78}. 
Sec.~\ref{Sec:Concl} contains a discussion of the results, their implications for the importance of the analogous
state-dependent quantum effects of the conformal anomaly in four dimensions, and outlook for the extension the results of
this paper to gravitational collapse in the full four-dimensional effective field theory (EFT) of gravity proposed in~\cite{EMEFT:2022}.

The paper also contains three appendices, wherein are collected for the convenience of the reader the curvature components 
in double null coordinates (Appendix \ref{App:Curv}), the metric functions for the collapsing null shell geometry (Appendix
\ref{App:Deriv}), and the stress tensors and horizon finiteness conditions in the various coordinates used, and relations 
between them (Appendix \ref{App:DNull}).

\section{Radial Collapse Geometry in Double Null Coordinates}
\label{Sec:Geom}

The general spherically symmetric line element in $3\!+\!1$ dimensions may be expressed in the factorized $2 \times 2$ form
\be
ds^2_4 = \g_{ab}\,dx^adx^b + r^2 d\W^2
\label{gensphsym}
\ee
where  $d\W^2 = d\th^2 + \sin^2\th\, d\f^2$ is the standard round line element on the
unit ${\mathbb S}^2$, $\g_{ab}(x^1, x^2)$ is the metric on the two-dimensional subspace 
of constant $\th, \f$, and $r= r(x^1, x^2)$ is a scalar function of the arbitrary two-dimensional coordinates 
$x^a \,(a=1,2)$. The radius $r$ is uniquely defined by the condition that the proper area of the 
sphere of constant $r$ is $A= 4 \pi r^2$ in the spherically symmetric spacetime.  

The various geometric quantities for the metric (\ref{gensphsym}) are given in Appendix \ref{App:Curv}. 
In particular the Einstein tensor of the full four-dimensional spacetime with the line element (\ref{gensphsym}) 
has the components~\cite{PoisIsrae:1988}
\vspace{-5mm}
\bes
\bea
&&G_{ab} =  \frac{ \g_{ab}}{r^2}\, \left[(\na r)^2 - 1 + 2\, r\sq r \right] 
- \frac{2}{r}\, \na_a \na_b r \,,\qquad a,b=1,2\\
&&\hspace{2cm} G^{\th}_{\ \th} = \,G^{\f}_{\ \f} = r \sq r - \frac{r^2}{2} R
\eea
\vspace{-2mm}
\label{Ein4}\ees
with all other components vanishing. In (\ref{Ein4}) we make use of the notations
\be
\na\!_a r = \pa_a r \equiv \frac{\pa r}{\pa x^a}\,,\qquad
(\na r)^2 \equiv \g^{ab}(\na_a r) (\na_b r)\,,\qquad \sq r \equiv \g^{ab} \na\!_a \na\!_b r
\label{dr}
\ee
with $\na\!_a$ the covariant derivative with respect to the two-dimensional metric $\g_{ab}$, and $R$ the corresponding 
two-dimensional Ricci scalar. We shall generally suppress any special notation distinguishing quantities derived from the 
two-dimensional metric $\g_{ab}$ {\it vs.}~the full four-dimensional line element (\ref{gensphsym}), as which is meant 
should be clear from the context. For example eqs.~(\ref{Ein4}) clearly refer to the four-dimensional Einstein tensor, 
since the Einstein tensor of any two-dimensional space vanishes identically. It is useful also to define the three functions 
$h, m$ and $\ka$ in terms of $r(x^1,x^2)$ by
\bes
\bea
&&h \equiv (\na r)^2  \equiv 1 - \frac{2Gm}{r} \label{defhm}\\
&&\ka \equiv - \frac{Gm}{r^2} = \frac{(\na r)^2 - 1}{2r}
\label{defkap}
\eea
\ees
which are also scalars with respect to the two-geometry $\g_{ab}$. The quantity $m$ is the 
Misner-Sharp mass function and $\kappa$ is the acceleration or surface gravity at $r$.\footnote{The definition
of $\kappa$ in this paper follows the conventions of \cite{PoisIsrae:1988}, which differ from the more general
definition of the surface gravity $\ka = \frac{1}{2}\sqrt{\frac{h}{f}} \frac{df}{dr}$. The two become equal,
except for a sign change, when $f=h$ and $m$ is independent of $r$.}

The Einstein eqs.~for the general spherically symmetric four-geometry (\ref{gensphsym}) are
\vspace{-2mm} 
\bes
\bea
&& - \na_a\na_b\, r + \big(\sq r + \ka\big)\, \g_{ab} = 4 \pi r \,G\,T_{ab}\\
&& \qquad \qquad \sq r - \frac{r}{2}\,R = 8 \pi r \,G\, p_{\perp}
\vspace{-3mm}
\eea
\label{genEin2}\ees

\vspace{-1.3cm}
\noindent
where
\be
T^{\th}_{\ \th} = T^{\f}_{\ \f} \equiv p_{\perp}
\ee
is the transverse pressure, which spherical symmetry requires must have equal $\th$ and $\f$ components.

If one defines the effective two-dimensional stress tensor $\t_{ab}$ by
\vspace{-1mm}
\be
T_{ab} \equiv \frac{\t_{ab}}{4 \pi r^2}\,,\qquad a,b = 1,2\,,
\label{t2D}
\vspace{-1mm}
\ee
covariant conservation of the full four-dimensional stress tensor gives~\cite{PoisIsrae:1988}
\vspace{-1mm}
\be
\na_b\t_a^{\ \, b} = 4\pi \na_ b\, \big(r^2 T_a^{\ \,b}\big) = 8\pi p_{\perp} \na_a r \,,
\label{Tcons}
\ee
all other components being satisfied identically. Hence the stress tensor $\t_{ab}$ is covariantly conserved 
purely in two dimensions {\it if and only if} the transverse pressure vanishes identically, {\it i.e.}
\vspace{-2mm} 
\be
\na_b \t_a^{\ \,b} = 0\,, \qquad \Leftrightarrow  \qquad p_{\perp}=0
\label{Ein2pb}
\ee
which we shall assume for a simplified model of gravitational collapse. This is a rather restrictive condition, about 
which we comment further in Secs.~\ref{Sec:2DAnom} and \ref{Sec:Concl}.

With the restriction $p_{\perp}\!=0$ the Einstein eqs.~(\ref{genEin2}) with (\ref{t2D}) become
\vspace{-3mm}
\bes
\bea
 - \na_a\na_b\, r + \big(\sq r + \kappa\big) \, \g_{ab}  &=& \frac{G}{r}\, \t_{ab}\,,\\
R = \frac{2}{r}\,\sq r 
\eea\label{Ein2pa}
\ees
which define a reduced 2D model, with a general covariantly conserved $\na_b \t_a\,^{\!b}= 0$.
By differentiating (\ref{defhm}) and using (\ref{defkap}) and (\ref{Ein2pa}) we obtain
the useful relation
\be
\frac{\pa m}{\pa x^a}= \big(\t_a^{\ \, b}  - \delta_a^{\ \,b}\, \t^{\ c}_c\big)\,\frac{\pa r}{\pa x^b}
\label{dma}
\ee
for the Misner-Sharp mass flux or gradient, where $\t^{\ c}_c= \g^{cd}\t_{cd}$ is the two-dimensional trace.

To this point the coordinates $(x^1,x^2)$ of the two-geometry at fixed $\th, \f$ have been left
arbitrary to emphasize covariance under arbitrary coordinate transformations of $(x^1,x^2)$.
We will make use of two specific useful choices of coordinates. The first is that of Schwarzschild coordinates,
obtained by identifying one of the coordinates ($x^2$ say) with $r$ itself. A possible  $dt\, dr$ cross term 
can be eliminated by a redefinition of $t$, so that $x^1$ can then be identified as the Schwarzschild time $t$.
This results in the line element taking on the standard Schwarzschild form~\cite{MTW}
\be
\g_{ab}\,dx^adx^b = -f\,dt^2 + \frac{dr^2\!}{\!h}
\label{Schw}
\ee
with $f$ and $h$ two functions of $(t,r)$. In these coordinates $h=g^{rr}$ is the same function defined in general 
two-dimensional coordinates by (\ref{defhm}), while (\ref{dma}) for $a=2, x^2=r$ becomes
\be
\frac{\pa m}{\pa r} = - \t^{\ t}_t= - 4\p r^2 \,T^{\ t}_t =  4\p r^2\r
\ee
in terms of the energy density $\r$. Integrating this eq.~with respect to $r$ shows that $m(t,r)$ is the 
Misner-Sharp mass-energy within the sphere of radius $r$ on the time slice fixed by $t$.

Since Schwarzschild coordinates (\ref{Schw}) become singular at $h=0$, and the causal structure 
is tied to the behavior of null rays, a different coordinate choice that proves useful is that of double 
null $(u,\vv)$ coordinates. These rely on the fact that every two-geometry is locally conformally flat, 
so the general two-dimensional line element (\ref{gensphsym}) can be expressed in the form
\vspace{-3mm}
\be
\g_{ab}\,dx^adx^b = - e^{2 \s} \, du\,d\vv
\label{doubnull}
\ee
with the metric $\g_{u\vv} = \g_{\vv u} = -\frac{1}{2} e^{2\s}$ and inverse $\g^{u\vv} =  \g^{\vv u} = -2 e^{-2\s}$, 
in terms of $\s (u,\vv)$. The line element (\ref{doubnull}) is invariant under the redefinitions
\vspace{-3mm}
\be
u \rightarrow \tilde u(u)\,,\qquad \vv \rightarrow \tv(\vv)
\label{dncoortrans}
\ee
\vspace{-2mm}
with the simultaneous redefinition of
\be
\s \rightarrow \tilde \s = \s  - \frac{1}{2} \ln \left(\frac{d\tilde u}{du}\right) 
- \frac{1}{2} \ln \left(\frac{d\tv}{d\vv}\right)\,,\qquad \frac{d\tilde u}{du}>0\,,\quad \frac{d\tv}{d\vv} >0\,.
\label{sigtrans}
\ee
Thus there is still considerable coordinate freedom to redefine $u$ and $\vv$ independently,
and we will make use of several different sets of double null coordinates. Since the conformal factor $e^\s$ 
changes under the coordinate transformation (\ref{dncoortrans})-(\ref{sigtrans}), such coordinate transformations 
are also conformal transformations, and form the infinite dimensional conformal group in two dimensions. The 
coordinate freedom can be fixed by {\it e.g.} setting $\s = 0$ in a region where the spacetime is flat, so 
that $u=t-r, \vv=t+r$ become the standard radial null coordinates in two-dimensional flat spacetime. 

In double null coordinates the coordinate invariant condition for the location of the apparent horizon (AH) is
\be
h = (\na r)^2 = - 4\,e^{-2 \s} \,\frac{\pa r}{\pa u} \frac{\pa r}{\pa \vv} 
\,\stackrel{AH}{=} \,0
\label{mots}
\ee
showing that the rate of change of the radius with respect to at least one of the null coordinates
must vanish there. The conditions
\vspace{-3mm}
\bes
\bea
&\ \ \,\displaystyle{\frac{\pa r}{\pa \vv}  = 0\qquad {\rm future\ AH}}\label{fmots}\\
&\displaystyle{\frac{\pa r}{\pa u}  = 0\qquad {\rm past\ AH}}\label{pmots}
\eea
\ees
define the future or past apparent horizons respectively, which are also invariant under (\ref{dncoortrans}).

The two-dimensional scalar curvature in double null coordinates (\ref{doubnull}) is
\be
R = -2 \sq \s = 8\, e^{-2\s}\,\frac{\pa^2\s}{\pa u\pa \vv}
\label{R2}
\ee
and the Einstein eqs. (\ref{Ein2pa}) with $p_{\perp}=0$ take the form of (\ref{Eindn}), 
which are covariant with respect to the two-dimensional coordinate/conformal transformation
(\ref{dncoortrans})-(\ref{sigtrans}). Thus $\t_{ab}\,dx^adx^b = \t_{\tilde a\tilde b}\, dx^{\tilde a}
dx^{\tilde b}$, so for example $\t_{uu}$ transforms as
\be
\t_{uu} = \left(\frac{d\tilde u}{du}\right)^2\,\t_{\tilde u \tilde u}
\ee
under (\ref{dncoortrans})-(\ref{sigtrans}). The Misner-Sharp mass is given by
\be
m(u,\vv) = \frac{r}{2G}
\left[1 + 4\, e^{-2\s}\, \left(\frac{\pa r}{\pa u}\right)\left(\frac{\pa r}{\pa \vv}\right)\right]\,,
\label{muv}
\ee
while eqs. (\ref{dma}) become
\vspace{-3mm}
\bes
\bea
&& \frac{\pa m}{\pa u} = 2\,e^{-2\s} \left(\t_{u\vv}\sdfrac{\pa r}{\pa u} 
-  \t_{uu} \sdfrac{\pa r}{\pa \vv}\right)\label{dmdu}\\[1ex]
&& \frac{\pa m}{\pa \vv} = 2\, e^{-2\s} \left(\t_{u\vv}\sdfrac{\pa r}{\pa \vv}
- \t_{\vv\vv}\sdfrac{\pa r}{\pa u}\right)\label{dmdv}
\eea
\label{dmdudv2}\ees
in double null coordinates.

\section{Classical Radial Collapse of a Null Shell}
\label{Sec:ClCollapse}

The simplest model of radial collapse which will form a BH classically is that of a spherical shell imploding 
upon its center at the speed of light. The classical energy-momentum-stress tensor 
of such a lightlike infalling shell is 
\be
\t_{\vv\vv}^{C} = \frac{dE}{d\vv}\,,
\label{tvvC}
\ee          
with $E(\vv)$ determining its profile as function of the advanced null coordinate time $\vv$, and with
all other components of $\t_{ab}^{C}$ vanishing. The total classical mass-energy
carried by the incoming null shell of radiation is
\be
M = \int_{-\infty}^{\infty}  \frac{dE}{d\vv}\, d\vv\,.
\label{Mdef}
\ee
The simplest case to analyze and solve explictly is that of an infinitesimally thin shell for which
\be
E(\vv) = M\, \th(\vv-\vv_0) \,,\qquad  \frac{dE}{d\vv} =   M\, \d(\vv-\vv_0)
\label{Edelta}
\ee
so that the four-dimensional classical energy-momentum tensor is
\be
T^C_{\vv\vv} = \frac{\t_{\vv\vv}^{C}}{4 \p r^2} = \frac{M}{4 \p r^2}\, \d(\vv-\vv_0)
\label{TvvC}
\ee
on the incoming null shell.

In this case the metric functions can be found explicitly in each region as follows.
In  the first region I, for $\vv<\vv_0$ interior to the imploding shell, spacetime is flat, so
that the two-dimensional line element at constant $\th, \f$ is
\bea
&&{\rm I:} \qquad ds^2= -dt^2 + dr^2 = -du\,d\vv  \,, \qquad {\rm with}\qquad u \equiv t-r, 
\quad \vv\equiv t + r  < \vv_0\,,\nn
&&\hspace{3.5cm} \s(u,\vv) = 0\,,\qquad r(u,\vv) = \frac{\vv-u}{2}
\label{regionI}
\eea
which satisfies (\ref{Eindn}) with $\t_{ab} =0$.

In the exterior region $\vv > \vv_0$ outside of the shell, the geometry is that of the
sourcefree four-dimensional Schwarzschild solution, {\it i.e.} the two-dimensional solution is
\bea
&& {\rm II}: \quad ds^2= f(r)\,\left(-dt^2 + dr^{*\,2}\right) = - f(r)\,d\tilde u\,d\tv  \,, \quad {\rm with}
\quad f(r) = 1 - \frac{\rM}{r}\,, \quad \rM \equiv \frac{2GM}{c^2} \nn
&& \qquad \quad dr^* = \frac{dr}{f(r)}\,,
\quad  r^* \equiv r + \rM \ln\left(\frac{r}{\rM} - 1\right)\,,
\qquad \tilde u \equiv t - r^*,\quad \tv \equiv t + r^* > \tv_0  \,.
\label{regionII}
\eea
We denote with tildes the Schwarzschild null coordinates $(\tu,\tv)$, since they are allowed to differ
from the corresponding $(u,\vv)$ coordinates in the flat region (\ref{regionI}).  The relations 
(\ref{regionII}) yield a solution to the sourcefree Einstein eqs. (\ref{Eindn}) with $\t_{ab} = 0$ and
\be
\tilde \s = \frac{1}{2} \ln f(r)\,,\qquad \frac{\tv - \tilde u}{2} = r^*= r + \rM \ln\left(\frac{r}{\rM} - 1\right)
\label{sigr2}
\ee
determining $r$ and $\tilde \s$ implicitly as functions of $\tv-\tu$, and $\tv + \tu = 2t$ in this Schwarzschild
region II.

The two sets of double null coordinates must be matched for a continuous ($\cC^0$) metric at $\vv=\vv_0$. 
This is accomplished by noting that the radius $r$ has the same invariant geometric meaning 
in terms of the four dimensional metric (\ref{gensphsym}) in either region. Comparison of (\ref{regionI}) 
and (\ref{sigr2}) shows that $\s \neq \tilde\s$, so that the solution in the two regions in these 
coordinates as they stand is discontinuous across the null shell. In order to find a solution to 
the geometry of the spherical collapse of a null shell with ${\cal C}^0$ continuous metric functions we 
utilize the gauge freedom (\ref{dncoortrans})-(\ref{sigtrans}) to match the solution I (\ref{regionI}) 
of the interior to the exterior solution II (\ref{regionII}).
 
For $r \gg \rM$ and $u, \tu \to -\infty$, both regions I and II are asymptotically flat, so that we may choose 
the advanced null coordinates $\vv$ and $\tv$ to be equal there. The reparametrization freedom in $\vv$ 
can be used to require the interior $\vv$ coordinate to match the exterior $\tv$ coordinate  for all $u, \tu$.
Hence
\be
\tv=\vv\,,\qquad d\tv = d\vv\,,\qquad \tv_0= \vv_0\,.
\label{vequal}
\ee
Then requiring the metric function $r= (\vv -u)/2$ from (\ref{regionI}) to be equal to that from (\ref{sigr2}) at the location
of the null shell at $\tv_0 = \vv_0$ gives 
\vspace{-2mm}
\be
r^*\big\vert_{\vv=\vv_0} = \frac{\vv_0 - \tilde u}{2} = r_0(u) +  \rM \ln\left(\frac{r_0(u)}{\rM} - 1\right)
\label{rstar}
\ee
with
\vspace{-3mm}
\be
r_0(u) \equiv r(u, \vv_0) = \frac{\vv_0 - u}{2}\,,
\label{r0def}
\ee
so that the radius $r$ is continuous across the shell. Eq.~(\ref{rstar}) determines~\cite{BalbFabPRD99}
\be
\tu(u) = u - 2r_{_M}  \ln\left(\frac{\vv_0 - u}{2r_{_M}} - 1\right)
\label{uutilde}
\ee
as a function of $u$ and
\vspace{-3mm}
\be
r^*(u,\vv) = r(u,\vv) + \rM \ln\left(\frac{r(u,\vv)}{\rM} - 1\right) 
= \frac{\vv -u}{2} + \rM \ln \left(\frac{r_0(u)}{\rM} - 1\right)
\label{rstaruv}
\ee
as an implicit function of the original $(u,\vv)$ of region I, in region II.

Differentiating (\ref{r0def}) and using $dr^* = dr/f(r)$, or directly from (\ref{uutilde}) we have
\be
\frac{d\tu}{du} = \frac{1}{f(r)}\bigg\vert_{r =r_0(u)}\! \equiv \frac{1}{f_0}=  \left( 1 - \frac{r_{_M}}{r_0(u)}\right)^{-1}  
=\left(1 - \frac{2\rM}{\vv_0 - u}\right)^{-1}
\label{dutildu}
\ee
so that using (\ref{sigtrans}) with (\ref{sigr2}) and (\ref{vequal}), we obtain 
\be
\s = \tilde \s + \frac{1}{2} \ln \left(\frac{d\tilde u}{du}\right) 
= \frac{1}{2} \ln \left(\frac{f(r)}{f(r_0)}\right) = \frac{1}{2} \ln \left(\frac{f}{f_0}\right)
\label{sigmatch}
\ee
in region II, determining also the second metric function $\s$ in the Schwarzschild region II,
now expressed in the original $(u,\vv)$ coordinates. Since (\ref{sigmatch}) vanishes at $\vv=\vv_0, r=r_0(u)$,
$\s (u,\vv_0)$ is continuous with $\s =0$, (\ref{regionI}) of the interior flat region I.  Thus the two-dimensional
line element
\be
ds^2 = -e^{2\s} \,du\,d\vv = - \frac{f(r)}{f(r_0)} \, du\, d\vv = -f(r) \, d\tu\,d\tv = -f(r)\, dt^2 + \frac{dr^2}{f(r)}
\ee 
is indeed the Schwarzschild exterior geometry in region II for $\tv =\vv>\vv_0$, after the passage of the null shell,
continuously matched to the flat region I at $\vv=\vv_0$, with the coordinate transformation (\ref{uutilde}).

The piecewise solutions to $r$ and $\s$ in the two regions and the full geometry determined by
the impolding null shell localized at $\vv=\vv_0$ according to (\ref{tvvC})-(\ref{Edelta}) can
be combined in terms of Heaviside step function 
\vspace{-2mm}
\be
\Theta(\vv-\vv_0) = \left\{ \begin{array}{lr} 1, \quad& \vv > \vv_0 \nn 0, \quad&\vv<\vv_0 \end{array} \right.
\ee
in the form
\be
\s (u,\vv) = \frac{1}{2} \ln\left(\frac{f(r)}{f(r_0)}\right)\, \Theta(\vv-\vv_0)
\label{sigsoln}
\ee
with $r(u,\vv)$ determined by the implicit relation for $\vv>\vv_0$ in region II
\be
r(u,\vv) = \frac{\vv-u}{2} +\rM \ln \left(\frac{r_0f_0}{rf}\right)\, \Theta(\vv-\vv_0)
=\frac{\vv-u}{2} +\rM\ln \left(\frac{r_0 -\rM}{r -\rM}\right)\, \Theta(\vv-\vv_0) 
\label{rsoln}
\ee
and $r_0(u)$ given by (\ref{r0def}). 

\begin{wrapfigure}{hr}{.44\textwidth}
\vspace{-1.2cm}
\includegraphics[height=10cm,width=7.2cm, trim=2cm 0cm 1.9cm 1.5cm, clip]{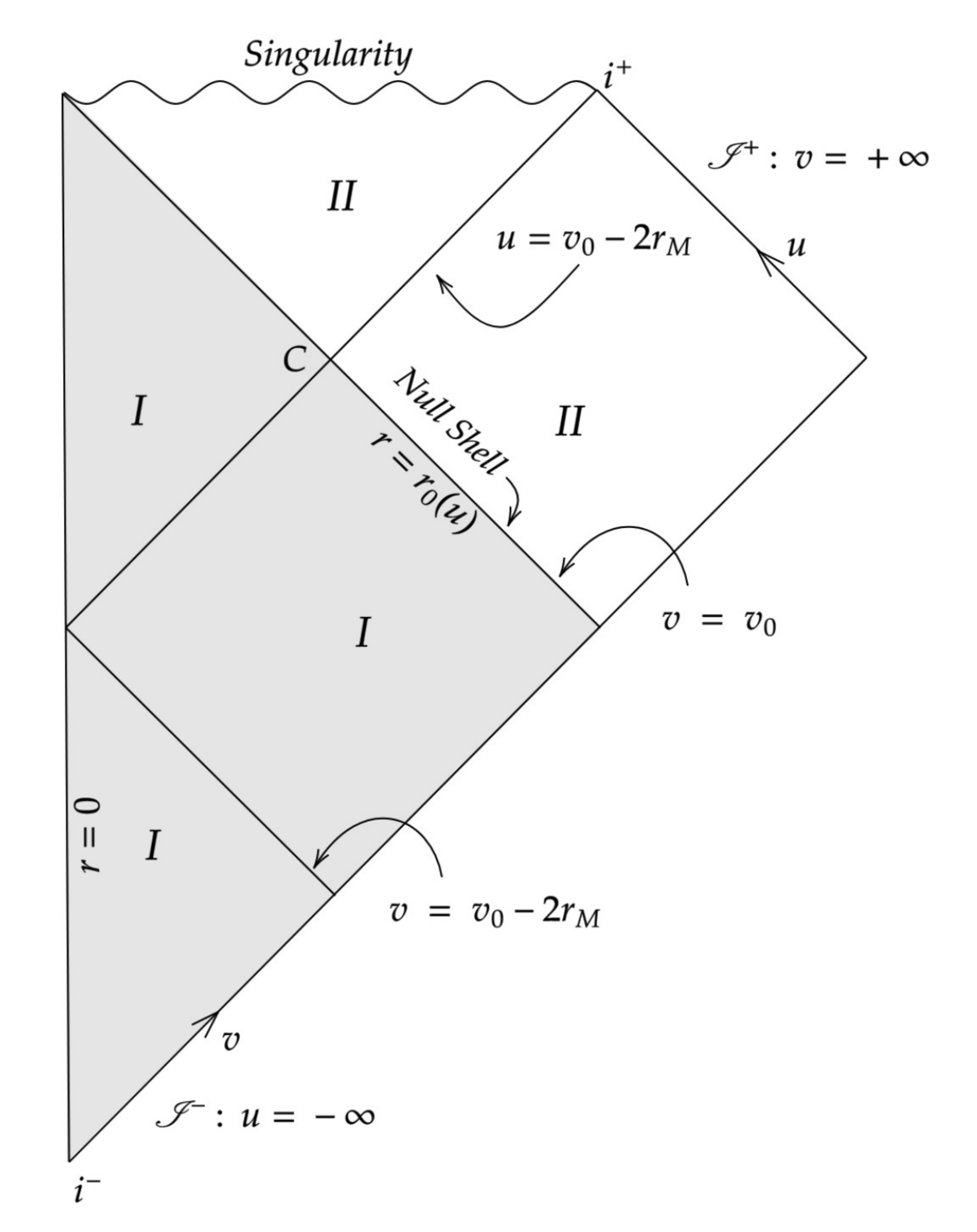}
\vspace{-7mm}
\singlespace
\caption{Carter-Penrose conformal diagram of radial collapse of a null shell. The shaded region I, $\vv<\vv_0$ is flat, 
while the unshaded region II, $\vv > \vv_0$ is Schwarzschild with mass $M$. The point C with coordinates (\ref{pointC}) 
is where the shell crosses its future event horizon. } 
\label{Fig:Penrose}
\vspace{3mm}
\end{wrapfigure}

From (\ref{sigsoln})-(\ref{rsoln}) it is clear that although $\s$ and $r$ are $\cC^0$ continuous at $\vv=\vv_0$, 
their first derivatives with respect to $\vv$ are not. Since the derivative of the Heaviside step function $\Theta$ is 
a Dirac $\d$-function, the second derivative 
\be
\frac{\pa^2 r}{\pa \vv^2} = -\frac{\rM}{2r} \, \d(\vv-\vv_0) + \dots
\label{rvvdelta}
\ee
contains a Dirac $\d$-function contribution at $\vv=\vv_0$ (with the ellipsis indicating the remaining terms
which are non-singular). The various first and second derivatives of $r$ and $\s$ with respect to $u$ and $\vv$ 
in each region are catalogued in Appendix \ref{App:Deriv}. With those full expressions
one may check that the classical Einstein eqs.~(\ref{Eindn}) are satisfied everywhere, including
the only component with a non-zero source
\vspace{-3mm}
\be
G_{\vv\vv} = 8 \p G \, T^C_{\vv\vv}
\label{TGvv}
\ee
from the stress tensor (\ref{TvvC}) of the null shell, with the $\d$-function from (\ref{rvvdelta}).
The Carter-Penrose conformal diagram for the classical geometry of the radially collapsing null shell of 
finite mass $M$ but infinitesimal thickness is illustrated in Fig.~\ref{Fig:Penrose}.

From (\ref{uutilde}) as the $u$ coordinate in region I approaches the finite value 
\vspace{-2mm}
\be
I: u \rightarrow \vv_0 - 2\rM
\vspace{-3mm} 
\ee
where $r \to \rM$, which corrresponds to
\vspace{-2mm}
\be
II: \tu \to +\infty\,,\quad \frac{\pa r}{\pa \vv} = \frac{f}{2} \to 0
\vspace{-1mm}
\ee
in the Schwarzschild region II, the condition (\ref{fmots}) is satisfied. Thus $u\!=\!\vv_0 - 2r_{_M}, \vv \! \ge \! \vv_0$
is the location of the future marginally outermost trapped surface and apparent horizon (AH). There is a 
last incoming null ray at $\vv\!=\!\vv_0- 2\rM$ which reflects from the origin at 
$u\!=\!\vv\!=\! \vv_0\!-\!2\rM$ and becomes the outgoing null ray defining the future BH horizon, but 
the conditions (\ref{mots})-(\ref{fmots}) are not satisfied until $\vv\! \ge\! \vv_0$. Incoming rays with 
$\vv_0 \!-\!2\rM\!< \!\vv\! <\! \vv_0$ reflect from the origin too late and are trapped, being pulled back 
finally to the future singularity at $r\!=\!0$. Thus the point $C$ at which 
the imploding null shell crosses its future horizon, with coordinates
\vspace{-2mm}
\be
(u,\vv)_C =(\vv_0 - 2\rM, \vv_0)
\label{pointC}
\vspace{-2mm}
\ee 
is where the AH and marginally trapped surface first appears, and a classical BH is formed, {\it cf.}~Fig.~\ref{Fig:Penrose}.

Since the approach of $u$ to the horizon is important in evaluating the quantum effects in the following
sections, we note that (\ref{rsoln}) may be written in the form
\be
\exp\left(\sdfrac{r}{\rM}\right)\, \left(\sdfrac{r}{\rM} - 1\right) = \exp\left(\sdfrac{\vv-u}{2\rM}\right) \,
\left( \sdfrac{r_0}{\rM}-1\right)\,,\qquad \vv > \vv_0\,,
\label{rexp}
\vspace{-1mm}
\ee
so that if $u = \vv_0 -2\rM (1+\e)$
\vspace{-3mm}
\be
\sdfrac{r_0}{\rM} = 1 + \e\,,\qquad \sdfrac{r}{\rM}  = 1 + \e \, \exp\left(\sdfrac{\vv-\vv_0}{2\rM}\right)  + \cO(\e^2)
\label{horizlimit}
\vspace{-1mm}
\ee
as $\e \to 0$. Thus both $r_0 \to \rM$ and $r \to \rM$ at fixed $\vv$ in the horizon limit, and both $f_0,f \to 0$,
while 
\vspace{-1mm}
\be
\sdfrac{f}{f_0} \to \exp\left(\sdfrac{\vv-\vv_0}{2\rM}\right)
\label{fratio}
\vspace{-2mm}
\ee
remains finite in this limit at fixed $\vv$ (while growing exponentially with $\vv$).

\section{The Stress Tensor of the Conformal Anomaly and the BH Horizon }
\label{Sec:2DAnom}

With the classical geometry of the imploding null shell forming a BH determined in Sec.~\ref{Sec:ClCollapse},
we turn to quantum effects in this two-dimensional spacetime. Since with $p_\perp \!= \!0$, $\t_{ab}$ is the conserved
stress tensor of the 2D spacetime at fixed $(\th,\f)$, we can model the quantum effects from the stress tensor of the 
two-dimensional conformal anomaly, which has been considered previously for the vacuum state in \cite{ParPirPRL94}.

We note in passing that the condition $p_\perp\!=\! 0$ does {\it not} follow from the dimensional reduction of the 4D theory to consideration
of the spherically symmetric $s$-waves only. Without the restriction $p_\perp\!=\! 0$ the $s$-wave reduction of the full 4D theory 
contains additional terms, as have been found and discussed in a number of papers~\cite{MukWipZel1994,BalbFabPRD99,BalbFabPLB99}. 
These additional terms in what is known as 2D dilaton gravity arise from the metric function $r(x^1,x^2)$ becoming a dilaton and an additional 
dynamical field in the effective 2D theory~\cite{CGHSPRD92,GruKumVas02}. However the 2D dilaton theory has been extensively 
studied and gives unphysical results for the 4D stress tensor in BH spacetimes, and for Hawking radiation in the gravitational collapse
problem~\cite{MukWipZel1994,BalbFabPRD99,BalbFabPLB99}. 

There are several reasons for this failure of the dimensionally reduced 2D dilaton theory to correctly reproduce even qualitatively
the features of the 4D theory, the principal one being the `dimensional reduction anomaly'~\cite{FroSutZel1999}. This is the fact 
that dimensional reduction does not commute with quantization and renormalization, since the 4D theory requires more
counterterms and counterterms of different types than the 2D theory. The result is that the $s$-wave contribution to the 
renormalized stress tensor of the 4D theory does {\it not} coincide with the renormalized stress tensor of the dimensionally
reduced 2D dilaton theory, which behaves in qualitatively different (and physically incorrect) ways from the 4D theory. 
For this reason the 2D dilaton theory of~\cite{MukWipZel1994,CGHSPRD92,GruKumVas02} is {\it not} the theory we consider 
or discuss in this paper. The true theory is intrinsically four dimensional, even in the case of spherical symmetry, and 
requires use of the four-dimensional conformal anomaly instead~\cite{BalbFabPRD99}. 

Since the 4D anomaly effective action and stress tensor is technically much more involved~\cite{EMVau:2006}, 
our purpose in this paper is to first study the state-dependent effects of the stress tensor derived from the 2D conformal 
anomaly on the future horizon in a simplified model of a 2D black hole, which requires that we impose the restriction
$p_\perp\!=\! 0$. 

In two dimensions the effective action corresponding to the conformal trace anomaly was given in Ref.~\cite{Poly:1987}
in the non-local form
\vspace{-2mm}
\be
S\!_{\rm anom}[\g]  = -\frac{N\hbar}{96\pi} \int\! d^2x\sqrt{-\g} \int\! d^2x'\sqrt{-\g'}\ 
R_x\,(\sq^{-1})_{x,x'} R_{x'}
\label{actPoly}
\vspace{-1mm}
\ee
where $N=N_s + N_f$ is the number of free massless fields (scalar or fermion) in the underlying QFT. 
This effective action is the result of functionally integrating out $N$ free massless quantum fields 
$\j_i, i= 1,\dots,N$ with classical action $S\!_{cl}[\j_i;\g]$ in two-dimensional curved spacetime, {\it i.e.}
\vspace{-1mm}
\be
\exp\left\{\sdfrac{i}{\hbar}\, S\!_{\rm anom}[\g] \right\} = 
\int \prod_{i=1}^N [\cD \j_i]\,  \exp\left\{\sdfrac{ i}{\hbar}\, S\!_{cl}[\j_i;\g]\right\} 
\label{fnint}
\vspace{-1mm}
\ee
which defines the one-particle irreducible (1PI) effective action of the quantum fields in a general 2D curved space
with metric $\g_{ab}$. The explicit factor of $\hbar$ in (\ref{actPoly}) reminds that this is the result of the quantum 
functional integral (\ref{fnint}). It gives the compact (and exact) result of all connected quantum one-loop stress tensor 
correlation functions $\lag \t_{a_1}^{\ b_1}(x_1) \dots  \t_{a_n}^{\ b_n}(x_n)\rag$ by successive variations
of $S\!_{\rm anom}[\g]$ with respect to the arbitrary metric $\g_{ab}$. A normalization factor, which drops out of 
all 1PI connected correlation functions for $n >1$ has been set equal to unity in (\ref{fnint}), so that $S\!_{\rm anom}[\g]$ 
and $\lag \t_{a}^{\ b}(x)\rag$ vanishes in infinite flat space with no boundaries. In other words, $S\!_{\rm anom}[\g]$ 
is the renormalized effective action functional, whose variations define the renormalized stress tensor correlation
functions, and no further renormalization is required. For the first variation we drop the brackets and write $\t_a^{\ b}$ for
$\lag \t_a^{\ b}\rag$.

In the form (\ref{actPoly}) it should be clear that non-local quantum effects are contained in this effective action through 
the boundary conditions needed to specify the Green's function $(\sq^{-1})_{x,x'}$ of the scalar wave operator. It is this 
essential non-local state dependence that leads to the possibility of novel quantum effects on BH horizons, which are not 
determined by the local curvature alone. However, the non-local action (\ref{actPoly}) may also be written in the local 
form 
\vspace{-1mm}
\be
S_{\!\!\cA}[\g;\vf]  \equiv -\frac{N\hbar}{96\pi} \int d^2x \sqrt{-\g}\,
\left(\g^{ab}\,\na\!_a \vf\,\na\!_b \vf - 2 R\,\vf\right)
\label{act2}
\vspace{-1mm}
\ee
by the introduction of a new scalar field $\vf$, called a {\it conformalon}, since shifts in $\vf$ correspond to conformal
transformations $e^\vf$ of the metric. The equivalence of (\ref{actPoly}) and (\ref{act2}) is demonstrated by variation 
of (\ref{act2}) with respect to $\vf$ which yields its eq. of motion
\vspace{-3mm}
\be
- \sq \vf = R
\label{eom2} 
\vspace{-2mm}
\ee
which is linear in $\vf$, since (\ref{act2}) is quadratic in $\vf$. If (\ref{eom2}) is formally solved for $\vf = -\sq^{-1} R$ 
by means of its Green's function, and substituted back into (\ref{act2}) the non-local form of the action (\ref{actPoly})
is recovered, up to a surface term. Clearly this inversion of (\ref{eom2}) is not unique since the Green's function $\sq^{-1}$
depends on as yet unspecified boundary conditions, which are in one-to-one correspondence with the specification
of the solution to (\ref{eom2}) by the fixing of solutions $\vf_0$ to the corresponding homogeneous eq.~$\sq\vf_0 =0$.
Thus in the local form (\ref{act2}), the state-dependent effects of the underlying QFT are contained in the  
choice of the particular homogeneous solution to the wave eq.~(\ref{eom2}). 

Varying the local form of the action (\ref{act2}) with respect to the two dimensional metric $\g^{ab}$ gives 
the energy-momentum tensor of the 2D quantum conformal anomaly
\be
\t_{ab}^{\cA} \equiv -\frac{2\!\!}{\!\!\sqrt{-\g}} \frac{\delta}{\d \g^{ab}}\,S_{\!\cA} [\g; \vf] 
= \frac{N\hbar}{48\pi}\left(2\na\!_a\na\!_b\vf - 2\g_{ab}\sq\vf + \na\!_a\vf\,\na\!_b\vf
- \sdfrac{1}{2} \g_{ab}\na\!_c\vf\na^c\vf\right)
\label{anomT2}
\ee
which is covariantly conserved in 2D, by use of (\ref{eom2}) and by virtue of the vanishing of the Einstein 
tensor in two dimensions. The trace of (\ref{anomT2}) reproduces the 2D trace anomaly \cite{BirDav}, {\it i.e.}
\vspace{-1mm}
\be
\t_a^{\cA \,a} = - \frac{N\hbar}{24\pi}\,\sq \vf = \frac{N\hbar}{24\pi} \,R
\label{Anom2D}
\vspace{-1mm}
\ee
upon making use of (\ref{eom2}). Henceforth we drop the superscript $\cA$ on the anomaly stress tensor
(\ref{anomT2}) to simplify notation, since it is clearly distinguished from the classical stress tensor $\t^C_{ab}$ of
the null shell in (\ref{tvvC})-(\ref{TvvC}).

\begin{figure}[ht] 
\vspace{-2mm}
\includegraphics[height=2.5cm, trim=0mm 0mm 0mm 0mm, clip=false]{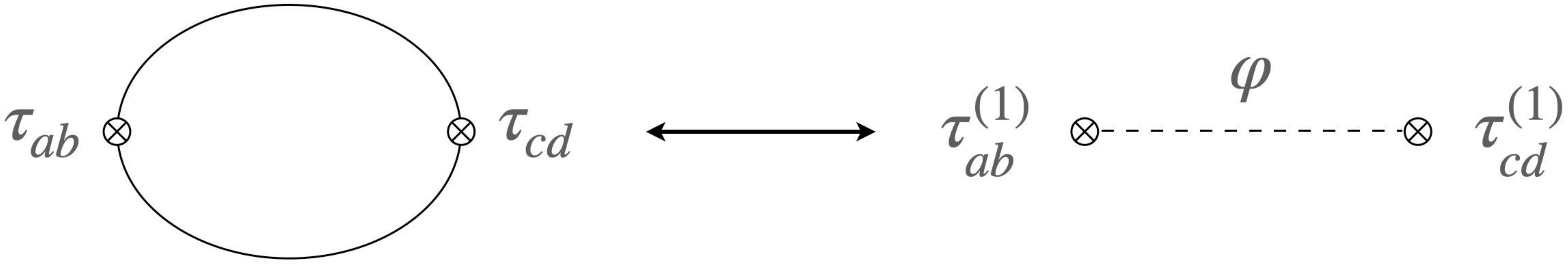}
\vspace{-2mm}
\caption{Left: The one-loop stress tensor vacuum polarization of a 2D CFT, which exhibits the massless $1/k^2$
pole of (\ref{Pi2D}). Right: The equivalent classical tree graph of the conformalon scalar $1/k^2$ propagator. See the text
and Ref.~\cite{BlaCabEM:2014} for the details of this correspondence.}
\vspace{-3mm}
\label{Fig:TT} 
\end{figure}

The scalar conformalon field $\vf$ may be regarded as an effective or collective degree of freedom that can be related 
to two-particle Cooper-pair intermediate states of the underlying massless conformal field 
theory~\cite{BlaCabEM:2014}. This may be seen by taking a second variation of (\ref{act2}) with respect to the 
arbitrary metric $\g^{cd}$ and then evaluating the result in flat space. This results in the vacuum polarization diagram 
of $\P_{abcd} =i \lag \t_{ab}\t_{cd}\rag $, whose intermediate two particle state exhibits a $1/k^2$ pole
in momentum space that can be expressed as the Greens' function propagator of the effective scalar 
degree of freedom $\vf$. Thus the one-loop $\P_{abcd}$ may be represented by a classical {\it tree}
graph in $\vf$, with no loops {\it cf.} Fig.~\ref{Fig:TT}. 

The one-loop polarization tensor in the underlying quantum theory has the form in momentum space
\vspace{-6mm}
\bes
\bea
&\displaystyle{\Pi_{abcd}(k)\Big\vert_{\rm 2D}  = \frac{N\hslash}{12\p k^2\!}\ 
\big(\h_{ab}k^2 - k_a k_b\big)\,\big(\h_{cd}k^2 - k_c k_d\big)} \label{Pi2D}\\
&\displaystyle{\Pi_{ab \ c}^{\ \ \,c}(k)\Big\vert_{\rm 2D}  = 
\frac{N\hslash}{12\p}\,  \big(\h_{ab}k^2 - k_a k_b\big)}\label{PiTr}
\vspace{-1mm}
\eea
\label{TT2D}
\ees
showing that the non-zero trace and coefficient on the right side of (\ref{Anom2D}) is directly related to 
the existence and residue of the $1/k^2$ pole in $\Pi_{abcd}$. In fact, once the tensor index structure 
indicated in (\ref{Pi2D}) is fixed, as required by symmetries and the covariant conservation Ward identity
$k^a\Pi_{abcd}(k)=0$ on any index, the one-loop diagram of Fig.~\ref{Fig:TT} is UV finite and 
completely determined, with (\ref{TT2D}) the result~\cite{BertKohl:AP01}. This shows that the conformal 
anomaly and pole is independent of the regularization scheme and detailed UV behavior of the quantum theory, 
provided that the identities following from the covariant conservation law (\ref{Tcons}) are maintained.

The correspondence with the propagator tree graph in Fig.~\ref{Fig:TT} is established by defining the vertex
$\t_{ab}^{(1)}$ by the term linear in $\vf$ in (\ref{anomT2}), {\it i.e.}
\vspace{-2mm}
\be 
\t_{ab}^{(1)} = \frac{N\hbar}{24\pi}\left(\na\!_a\na\!_b\vf - \g_{ab}\sq\vf\right)
\label{tau1}
\ee
and recognizing that the normalization of the $\vf$ field in (\ref{act2}) differs by a factor of $N\hbar/48\p$
from that of a canonically normalized scalar field, so that its propagator is $(48 \p/N\hbar) \times 1/k^2$.
Attaching the vertex factor (\ref{tau1}) to each vertex in the $\vf$ tree graph of Fig.~\ref{Fig:TT} 
and taking account of the normalization of the $\vf$ propagator gives for the $\vf$ tree graph in momentum space
\be
\left(\frac{N\hbar}{24\pi}\right)^2\,\left(\frac{48 \p}{N \hbar \,k^2}\right)\big(\h_{ab}k^2 - k_a k_b\big)\,\big(\h_{cd}k^2 - k_c k_d\big)
=  \frac{N\hslash}{12\p k^2\!} \ \big(\h_{ab}k^2 - k_a k_b\big)\,\big(\h_{cd}k^2 - k_c k_d\big)
\label{phiT2}
\ee
which coincides with (\ref{Pi2D}), establishing their equivalence. Note that the classical theory of 2D gravity defined by 
$\int d^2x\! \sqrt{\g}\, R$ has no transverse modes and no propagating degrees of freedom at all, so the $1/k^2$ propagator 
and effective scalar  degree of freedom it describes arises entirely from the quantum effect of the anomaly, described by (\ref{act2})
in which $\hbar$ is a parameter, but in terms of an effective classical field satisfying (\ref{eom2})~\cite{EMZak:2010,BlaCabEM:2014}.

The essential point now is that the massless pole in (\ref{Pi2D}), equivalently (\ref{phiT2}), is a {\it lightlike} singularity, 
signaling significant effects of the quantum conformal anomaly on the light cone, which extends to macroscopic distance 
scales, irrespective of the local curvature $R$. To see the effect of the anomaly and $\vf$ on horizons directly, and to relate 
it to the classical BH geometry of the Sec.~\ref{Sec:ClCollapse}, consider the 2D line element of the Schwarzschild form (\ref{regionII}).
The components of the 2D anomaly stress tensor (\ref{anomT2}) in the $(t,r^*)$ coordinates of (\ref{regionII}) are
\begin{subequations}
\begin{align}
\t^{\ \,t}_t &= \frac{N\hbar}{24\p}\left\{ - \frac{1}{4f} \left( \dot\vf^2 + \vf_{,r^*}^2 - 2 f'\vf_{,r^*}\right) 
- \frac{\ddot\vf}{f} +R\right\}\\
\t^{\ \,t}_{r^*}& = \frac{N\hbar}{48\p f}\, \Big\{-\!2\,\dot\vf_{,r^*} + \dot\vf\, \big(f' - \vf_{,r^*}\big)\,\Big\}\\
\t^{\ \,r^*}_{r^*}& =  \frac{N\hbar}{24\p} \left\{\frac{1}{4f} \left( \dot\vf^2 + \vf_{,r^*}^2 - 2 f'\vf_{,r^*}\right) 
+ \frac{\vf_{,r^* r^*}}{f} +R\right\}
\end{align}
\label{T2Dtrstar}
\end{subequations}
where $\vf_{,r^*} = \frac{\pa \vf}{\pa r^*\!}$ and $\vf_{,r^* r^*} = \frac{\pa^2 \vf}{\,\pa r^{*\,2}\!}\,$.

The linear eq.~(\ref{eom2}) for $\vf$ is
\be
\sq \vf = - \sdfrac{1}{f} \sdfrac{\pa^2\vf }{\pa t^2} +  \sdfrac{\pa}{\pa r} \left(f\,  \sdfrac{\pa \vf}{\pa r}\right)
= \sdfrac{1}{f} \left(- \sdfrac{\pa^2 }{\pa t^2} +  \sdfrac{\pa^2}{\pa r^{*\,2}}\right)\vf = -R = f'' = \sdfrac{d^2f}{dr^2}
\label{eomtx}
\ee
in these coordinates. A particular solution to this inhomogeneous eq.~is $\vf =\ln f$. The associated homogeneous wave 
eq.~has general wave solutions $\exp\{ik (r^* \pm t)\}$. If we are interested in stationary states, and restrict 
to $k=0$, we may illustrate the behavior of the anomaly stress tensor on the horizon with linear functions of $t$ and $r^*$. 
In this case one can examine the effect of a stationary state solution of (\ref{eomtx}) in the form
\vspace{-2mm}
\be
\vf_{_{P,Q}}= Pt + Q r^*  + \ln f(r) = \sdfrac{P+Q}{2}\, \vv + \sdfrac{P-Q}{2}\, \tu + \ln f(r)
\label{phisoln}
\ee
where an irrelevant constant is set to zero because (\ref{act2}) and (\ref{anomT2}) depend only upon the
derivatives of $\vf$. Substituting this solution into the stress tensor (\ref{eom2}) with $\vf_{,r^*} = Q + f'$
and $\vf_{,r^* r^*} = ff''$, we find
\begin{subequations}
\begin{align}
\t^{\ \,t}_t &= -\frac{N\hbar}{24\pi}\left\{\frac{1}{4f} \left(P^2 + Q^2 - f^{\prime\,2}\right) + f''\right\}\\
\t^{\ \,t}_{r^*}& = -\frac{N\hbar}{48\pi} \frac{PQ}{f}\\
\t^{\ \,r^*}_{r^*}& =  \frac{N\hbar}{96\pi} \frac{1}{f} \left(P^2 + Q^2 - f^{\prime\,2}\right) 
\end{align}
\label{T2Dstat}\end{subequations}
in the $(t,r^*)$ coordinates. If one then specializes to the Schwarzschild exterior line element of (\ref{regionII}), with 
\be
f(r) = 1 - \sdfrac{\rM}{r} \,,\qquad f' = \sdfrac{\rM}{r^2}\,,\qquad f''=  -\sdfrac{2\rM}{r^3}=-R
\ee
the stress tensor (\ref{T2Dstat}) of the quantum anomaly becomes
\begin{subequations}
\begin{align}
\t^{\ \,t}_t &= -\frac{N\hbar}{24\pi}
\left\{\frac{1}{4f} \left(\frac{p^2 + q^2}{\rM^2} - \sdfrac{\rM^2}{r^4}\right) -  \sdfrac{2\rM}{r^3} \right\}\\
\t^{\ \,t}_{r^*}& = -\frac{N\hbar}{48\pi \rM^2} \frac{pq}{f}\\
\t^{\ \,r^*}_{r^*}& =  \frac{N\hbar}{96\pi} \frac{1}{f} \left(\frac{p^2 + q^2}{\rM^2} - \sdfrac{\rM^2}{r^4}\right) 
\end{align}
\label{T2DSch}\end{subequations}
where we have set the constants $P=p/\rM$ and $Q=q/\rM$, so that $(p,q)$ are dimensionless.

Eqs.~(\ref{T2DSch}) show that the stress tensor due to the quantum anomaly generically gives {\it divergent} $1/f$ 
contributions as $r\!\to\!\rM, f\!\to\!0$ on the BH horizon, irrespective of the small curvature there. This is a reflection of the
$1/k^2$ light cone singularity of (\ref{Pi2D}). The divergences can be arranged to cancel 
on the future horizon by the particular choice $p\!=\!-q =\!\pm 1/2$, or on the past horizon
by the choice $p\!=\!q =\pm 1/2$, corresponding to the future or past Unruh states~\cite{Unruh:1976}, or 
on both horizons by the choice $p\!=\!0, q\!=\! \pm 1$, corresponding to the Hartle-Hawking thermal 
state~\cite{HartHawk:1976,Israel:1976,GibPerry:1976} at the price of being non-vanishing as $r\to \infty$
(and being thermodynamically unstable due to negative heat capacity~\cite{HawkSpecHeat:1976}).

Any other values for $(p,q)$ result in divergences on the horizon. If one requires a time independent 
truly static solution then $p=0$. The case $p=q=0$ is both time independent and gives a $\vf$ and
stress tensor that tends to zero as $r\to \infty$, corresponding to asymptotically flat
conditions, but for this choice
\be
\t^{\ \, b}_{a}\big\vert_{p=q=0} \to -\frac{N\hslash}{96\p \rM^2 f}\ \left(\,\begin{array}{cl}\!\!\!-1\ & 0\\
\,0& 1 \end{array}\right) \to \infty \qquad {\rm as} \qquad r \to \rM
\label{Thoriz2D}
\ee
which diverges on the two-dimensional horizon as $r\! \to\!\rM, f\! \to\! 0$. These conditions correspond to the Boulware 
state~\cite{Boul:1975,ChrFul:1977}. 

The significance of the solution $\vf \!=\!\ln f$ to (\ref{phisoln}) corresponding to this state is
that $e^\vf \!=\! f$ is the conformal transformation that takes the 2D flat line element $-dt^2 + dr^{*\,2}$ to the
curved space line element of (\ref{regionII}). The stress tensor (\ref{Thoriz2D}) is the effect on the expectation
value of $\t_a^{\ b}$ of this conformal transformation on the quantum vacuum state. In this way the local 
conformalon scalar incorporates information about the non-local quantum state over the entire $t=const.$
Cauchy surface, relating the value of $ \t_a^{\ b}$ to the standard Minkowski vacuum state in the 
asymptotically flat region where $f\!\to\!1$ and $\vf \!\to\! 0$. The divergence of $\vf=\ln f$ as $r\!\to\!\rM$ reflects
the vanishing of the norm of the timelike Killing vector $\pa_t$ on the horizon, and breakdown of the separation
of positive and negative frequency (particle and anti-particle) solutions of the underlying quantum field theory,
upon which the definition of the unique quantum vacuum state in flat Minkowski space is based.

The results (\ref{T2DSch}) show that the special states which are regular on the horizon are isolated points 
of measure zero in the two-parameter space of general $(p,q)$, and in particular, there is no value of $(p,q)$ 
which yields a time independent regular solution for $\vf$ and (\ref{T2DSch}) on both the horizon and as 
$r \to \infty$. Apart from these specific states and particular values of $(p,q)$, each 
of which would require a rather technically involved calculation and renormalization of a quantum stress tensor 
to derive directly from the underlying quantum field theory in curved space, the effective action (\ref{act2}) 
of the conformal anomaly and its stress tensor (\ref{anomT2}) permits consideration of a wide class of non-vacuum 
initial states and their possible quantum effects, simply by changing the integration constants or more general 
homogeneous solutions of the conformalon $\vf$ field eq.~(\ref{eom2}). This permits the investigation of quantum 
effects of non-vacuum initial conditions for general quantum fields on the BH horizon very simply and systematically.

\section{Non-Vacuum Initial States and Suppression of the Hawking Flux}
\label{Sec:Init}

To apply the anomaly stress tensor (\ref{anomT2}), (\ref{T2Dtrstar}) for non-vacuum states in the case
of gravitational collapse of the null shell and formation of the BH considered in Sec.~\ref{Sec:ClCollapse}, consider
eq.~(\ref{eom2}) in the double null coordinates (\ref{doubnull}) 
\vspace{-2mm}
\be
\frac{\pa^2 \vf}{\pa u \pa \vv} = 2\, \frac{\pa^2 \s}{\pa u \pa \vv}
\label{phieomdn}
\ee
the general solution of which may be expressed
\vspace{-2mm}
\be
\vf (u,\vv) = 2\, \Big[ \s(u,\vv) +  A(u) + B(\vv)\Big]
\label{genphisoln}
\vspace{-2mm}
\ee
in terms of two arbitrary functions $A(u), B(\vv)$. The particular solution $\vf=2 \s$ with $A\!=\!B\!=\!0$ gives 
$\t_{ab} = 0$ in the flat region I, corresponding to the initial state being the Minkowski vacuum. 
However in the Schwarzschild region II, $\vf= 2\s= \ln (f/f_0)$ from (\ref{sigmatch}). Note that in relation 
to (\ref{phisoln}), $\vf = \ln f - \ln f_0$ in region II corresponds to adding a particular homogeneous solution,
namely $-\ln f_0(u)$ to the solution of the inhomogeneous eq., $\ln f$. Tying $\vf$ rigidly to the geometry 
in this way, with a very particular homogeneous solution to the $\vf$ eq.~(\ref{phisoln}), as was assumed 
in earlier works \cite{Unruh:1976,DavFulUnrPRD76,ParPirPRL94} corresponds to the Unruh vacuum initial 
conditions after the passage of the null shell in the Schwarzschild region II, as we shall see presently.

The formulation in terms of a local independent field $\vf$ is considerably more general and allows for 
arbitrary homogeneous solutions of the differential eq. (\ref{eom2}) to be added as in (\ref{genphisoln}), 
corresponding to non-vacuum initial states. Substituting the general solution (\ref{genphisoln}) for $\vf$ 
into the stress tensor (\ref{anomT2}) we obtain the general form of the two-dimensional quantum anomaly 
stress tensor in the double null coordinates, with components
\vspace{-4mm}
\bes
\bea
&&\t_{uu} = \frac{N\hbar}{12\pi}\left[ \frac{\pa^2 \s}{\pa u^2} 
- \left(\frac{\pa \s}{\pa u}\right)^2
+ \frac{d^2 A}{du^2} +  \left( \frac{d A}{du} \right)^2 \right]\\
&&\t_{u\vv} = -\frac{N\hbar}{12\pi} \frac{\pa^2 \s}{\pa u\pa \vv}\,,\\
&&\t_{\vv\vv} = \frac{N\hbar}{12\pi}\left[ \frac{\pa^2 \s}{\pa \vv^2} 
- \left(\frac{\pa \s}{\pa \vv}\right)^2
+ \frac{d^2 B}{d\vv^2} + \left( \frac{d B}{d\vv} \right)^2 \right].
\eea\label{Tuv}\ees
It should be noted that (\ref{Tuv}) does not obey classical positivity conditions, nor should that
be expected for the expectation value of a quantum stress tensor \cite{BirDav}.

In the Schwarzschild region II (\ref{Tuv}) may be evaluated in the classical background geometry ({\it i.e.} ignoring
backreaction), with the aid of eqs.~(\ref{derivsig}) to obtain
\bes
\bea
&&\t_{uu} = \frac{N\hbar \rM}{48\pi f_0^2}\left[ \frac{1}{r_0^3} - \frac{1}{r^3}
+ \frac{3\rM}{4\ } \left(\frac{1}{r^4} - \frac{1}{r_0^4}\right)\right] 
+  \frac{N\hbar}{12\pi}\left[\frac{d^2 A}{du^2} + \left( \frac{d A}{du} \right)^2 \right]\,, \label{tauQuu}\\
&&\t_{u\vv} = -\frac{N\hbar \rM}{48\p r^3} \frac{f}{f_0}\,\\
&&\t_{\vv\vv} = -\frac{N\hbar \rM}{48\p r^3}\left(1- \frac{3\rM}{4 r\ }\right)
+ \frac{N\hbar}{12\pi} \left[\frac{d^2 B}{d\vv^2} + \left( \frac{d B}{d\vv} \right)^2 \right]\,.
\label{tauQvv}
\eea
\label{TUVsig}\ees
for $\vv> \vv_0$. An important observation about the vacuum $A\!=\!B\!=\!0$ terms in (\ref{TUVsig}) is that all 
components satisfy the finiteness conditions of \cite{ChrFul:1977} and Appendix \ref{App:DNull}. In particular, although
$\t_{uu}$ of (\ref{tauQuu}) contains a factor of $1/f_0^2$, the quantity in square brackets multiplying
it vanishes up to second order in $\e$ in the expansion near horizon limit (\ref{horizlimit}).

From the last eq.~(\ref{tauQvv}) for $\t_{\vv\vv}$ it is also clear that the function $B(\vv)$ adds to the classical stress tensor 
of the null shell (\ref{TvvC}) an ingoing flux contribution from non-vacuum initial conditions at $\sI^-$, which 
would change the mass $M$ and position of the BH horizon, but is otherwise of no particular interest for the 
behavior of the geometry near the future horizon, or the Hawking effect on $\sI^+$. Therefore we set 
$B(\vv) \!=\!0$ and focus on the possible effects of non-vacuum initial conditions determined by $A(u)$.

Evaluating the derivatives of the flux of energy associated with the quantum energy-momentum 
tensor (\ref{TUVsig}) with $B\!=\!0$, from the time derivative of the Misner-Sharp mass 
in region II in the Schwarzschild $(t,r)$ coordinates using (\ref{dma}) we find
\bea
&&\frac{\pa m}{\!\pa t}\,\bigg\vert_{B=0}  =  f_0\,\frac{\pa m}{\pa u}
+  \frac{\pa m}{\pa \vv} = - f_0^2\, \t_{uu} + \t_{\vv\vv}\nn
&& = -\frac{N\hbar r_{_M}}{48\pi r_0^3}\left(1 -  \frac{3r_{_M}}{4r_0\, }\right)
-  \frac{N\hbar f_0^2}{12\pi}\left[\frac{d^2 A}{du^2} + \left( \frac{d A}{du} \right)^2 \right]\,.
\label{dmdt}
\eea
For the vacuum initial conditions, $A\!=\!B\!=\!0$, at late times $t\to \infty$ as $\tu, \vv \to \infty, u\to \vv_0-2\rM$
at future null infinity $\sI^+$, $r_0 \to \rM$ and outgoing quantum energy flux goes to the limit
\be
\dot m_{_H} = \frac{\pa m}{\!\pa t}\,\bigg\vert_{A=B=0} \to -\frac{N\hbar}{192\pi r_{_M}^2} =  
- \frac{N\pi}{12\hbar}\,  (k_{_B}T_{_H})^2 
\label{Haw2D}
\ee
which is exactly the flux of $N$ quantum fields radiating at the Hawking temperature $T_{\!_H} = \hbar/(8 \pi k_{_B} GM)$
in two dimensions expected in the Unruh state. We obtain the Hawking flux for two dimensions and not four 
dimensions because we are using the two-dimensional conformal anomaly as a proxy for the quantum anomaly 
in four dimensions. This is in agreement with earlier results~\cite{Unruh:1976,DavFulUnrPRD76,ChrFul:1977,ParPirPRL94}. 

Note that the full energy flux (\ref{dmdt}) is a function only of $u$ if $B=0$ (as we neglect any backreaction)
and that the factor of $f_0^2$ multiplying $\t_{uu}$ can lead to a finite result at late times on 
$\sI^+$ as $u\to \vv_0-2\rM, f_0 \to 0$, only if there is a compensating factor of $1/f_0^2$ in (\ref{tauQuu}).  
Stated in a different way, the Hawking flux result (\ref{Haw2D}) is dependent upon the regularity of the
vacuum stress tensor on the horizon, but conversely if the regularity conditions are violated by non-vacuum
terms from $A(u)$, then they can change the energy flux (\ref{dmdt}) at $\sI^+$ at late times.
This is possible if and only if the non-vacuum terms in $\t_{uu}$ are $1/f_0^2$ singular on the 
future horizon, consistent with the analysis of~\cite{FredHag:1990}.

Comparing the general solution (\ref{genphisoln}) for $\vf$ in the Schwarzschild region II after the null shell
collapse with the particular solution (\ref{phisoln}) in the static Schwarzschild geometry,
we see that it corresponds to $p=-q$ and
\vspace{-3mm}
\be
A(u)\big\vert_{p=-q} =  \left(q + \sdfrac{1}{2}\right) \ln f_0 + A_{\rm reg}(u)\,,\qquad \text{where} \qquad
A_{\rm reg}(u) = -\sdfrac{\!q u}{2\rM} + q \ln \left(\sdfrac{r_0}{\rM}\right) 
\label{Aq}
\ee
and the latter $A_{\rm reg}(u)$ is finite and regular on the horizon, $u= \vv_0 - 2\rM, r_0 =\rM$. Since the important
effects on the horizon are associated with the divergent $\ln f_0$ term, we drop the regular contributions
and consider the effects of the simpler non-vacuum perturbation of the form
\vspace{-2mm}
\be
A(u) = \left(q + \sdfrac{1}{2}\right) \,\ln f_0 = \left(q + \sdfrac{1}{2}\right) \,\ln \left( 1 - \sdfrac{\rM}{r_0}\right)\,,
\qquad r_0 > \rM\,.
\label{Au}
\ee
This gives the additional contribution to $\t_{uu}$ 
\begin{align}
\t^A_{uu} = \frac{N\hbar}{12 \p} \left[\frac{d^2A}{du^2} + \left(\frac{dA}{du}\right)^2\right]
= \frac{N\hbar}{48 \p}\left(q^2 -\sdfrac{1}{4}\right) \frac{\rM^2}{r_0^4 f_0^2}
- \frac{N\hbar}{24 \p}\left(q + \sdfrac{1}{2}\right) \frac{\rM}{r_0^3 f_0}
\label{tauA}
\end{align}
in (\ref{tauQuu}), which has the $1/f_0^2$ behavior in the horizon limit $f_0\!\to \!0$ required to give 
a non-vanishing contribution to the flux (\ref{dmdt}) at late times. Thus we now find
\begin{align}
\frac{\pa m}{\pa t}&= \frac{N\hbar}{48 \p}\left[ -\frac{\rM}{r_0^3} + \frac{3 \rM ^2}{4r_0^4}
-\left(q^2 - \sdfrac{1}{4}\right) \frac{ \rM ^2}{r_0^4}  + \left(q + \sdfrac{1}{2}\right) \frac{2\rM}{r_0^3}\,f_0 \right]\nn
&\to -\frac{N\hbar}{48 \p \rM^2}\,q^2
\end{align}
as $u\!\to\! \vv_0-2\rM, r_0\!\to\! \rM, f_0\!\to\! 0$ at late times. If $q\!= \!- 1/2$ and the non-vacuum perturbation
(\ref{Au}) vanishes, one recovers the Hawking vacuum flux (\ref{Haw2D}) in the Unruh state, which is regular
on the future horizon, but if $q=0$ this flux is precisely cancelled, corresponding to the Boulware state, which
has a singular stress tensor (\ref{Thoriz2D}) on the horizon, and there is no Hawking radiation.

It is clear from this exercise that the Hawking flux and the behavior of the stress tensor on the horizon
are intimately connected and dependent upon one another, and both are determined by the particular
solution of the $\vf$ eq.~(\ref{eom2}) and stress tensor (\ref{anomT2}). That the assumption of regularity 
of the stress tensor on the horizon implies the Hawking effect was shown in Ref.~\cite{FredHag:1990}. 
The considerations above show that the converse is also true, namely a singular contribution to the quantum 
stress tensor $\t_{uu}$ from an initial state perturbation can modify or even eliminate the Hawking flux. 

Now a strictly divergent perturbation is disallowed by the requirement that the initial state be UV finite with
a Hadamard two-point function in QFT, in accordance with a theorem of \cite{FulSweWald:CMP78}. 
Any $A(u)$ homogeneous solution to (\ref{eom2}), if followed backwards in time and reflected from the 
origin must have been present in the initial state as incoming radiation in $B(\vv)$. Hence requiring that $B(\vv)$ 
be non-singular in the initial state on $\sI^-$ prior to collapse implies that $A(u)$ must also be non-singular 
on the horizon, and the strictly diverging behavior of (\ref{Au}) on the future horizon in (\ref{tauA}) is excluded. 

On the other hand, there is no need for the quantum stress tensor to diverge. If it becomes
arbitrarily large, while still finite, it can produce backreaction effects on the horizon that could lead to significantly 
different results than those obtained with vacuum initial data. Quantitative control of this large growth
of the stress tensor on the horizon requires regulating the logarithmic divergence of (\ref{Au}) and the
corresponding $1/f_0^2$ divergence of (\ref{tauA}) by a smooth cutoff for small but finite $f_0$.

\begin{wrapfigure}{hr}{.52\textwidth}
\vspace{-1cm}
\includegraphics[height=5.6cm, trim=1mm 2mm 2mm 8mm, clip=true]{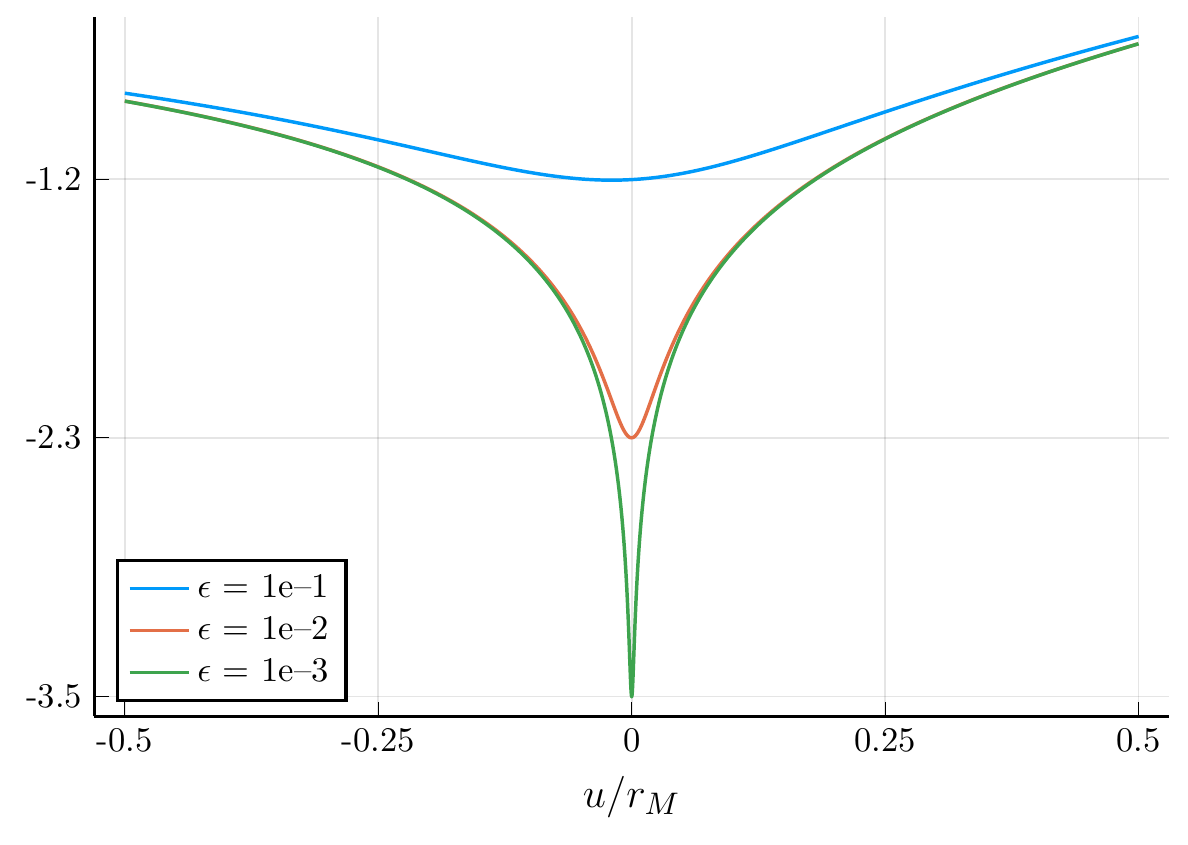}
\vspace{-6mm}
\singlespace
\caption{The regularized perturbation in the initial conditions (\ref{Aeu}) for $q=0$ and various $\e$.}
\label{Fig:APert} 
\end{wrapfigure}

Let the divergence in the $\t_{uu}$ component of the stress tensor in the non-vacuum state described 
by (\ref{Au}) be regulated by a small quantity $\e \ll 1$, such that (\ref{Au}) holds nearly everwhere but as 
$f_0 \to 0$, the logarithm is cut off by $\e$. That is, let $A(u)$ of (\ref{Au}) be replaced by $A_\e(u)$ such that
\vspace{-3mm}
\be
\lim_{\e \to 0^+} A_{\e}(u) = \left(q + \sdfrac{1}{2}\right) \,\ln |f_0| 
\label{eps0}
\vspace{-2mm}
\ee
but also such that
\vspace{-3mm}
\be
\lim_{u \to \vv_0-2\rM} A_{\e}(u) \to \left(q + \sdfrac{1}{2}\right)\,  \ln \e 
\label{logeps}
\ee
remains finite, regulated by the non-zero value of $\e\ll 1$. One simple such regulated $A(u)$
(by no means unique), with the required properties in the near horizon region might be
\be
\tilde A_{\e}(u) = \sdfrac{1}{2} \,\left(q + \sdfrac{1}{2}\right)\,  \ln \left(f_0^2 + \e^2\right) =
 \sdfrac{1}{2} \,\left(q + \sdfrac{1}{2}\right)\,  \ln \left[ \left(1 - \sdfrac{\rM}{r_0(u)}\right)^2 + \e^2\right]
\label{Aeu1}
\ee
which unlike (\ref{Au}) is also defined for $f_0 <0$. We may also require that $A_\e(u)$ have no singular behavior 
at any other $u$, whereas (\ref{Aeu}) still exhibits singular behavior at the origin $u=\vv_0,r_0 =0$ where $f_0 \to -\infty$.
Thus another possible fully regularized $A(u)$ is
\vspace{-2mm}
\be
A_{\e}(u) = \sdfrac{1}{2} \,\left(q + \sdfrac{1}{2}\right)\, \left\{
 \ln \left[ \left(\sdfrac{r_0(u)}{\rM} - 1\right)^2 + \e^2\right] - \ln \left[ \left(\sdfrac{r_0(u)}{\rM}\right)^2 + \e^2\right]\right\}
\label{Aeu}
\vspace{-1mm}
\ee
where both logarithmic singularities of (\ref{Au}) at $r_0 = \rM$ and $r_0 = 0$ are removed and 
regularized by the same $\e \ll 1$ small parameter. Then 
\vspace{-6mm}
\be
A_{\e}(u) \to \pm \left(q + \sdfrac{1}{2}\right)\,\ln \e
\vspace{-4mm}
\ee
for $u \to \vv_0 - 2 \rM$ or $u \to \vv_0$, respectively, as $\e \to 0^+$. This regularized function $A_\e(u)$ is
shown as a function of $u$ for $q=0$ and various $\e$ in Fig.~\ref{Fig:APert}.

\begin{figure}[h] 
\vspace{-4mm}
\begin{center}
\includegraphics[height=6.2cm, trim=6mm 8mm 0mm 2mm, clip=false]{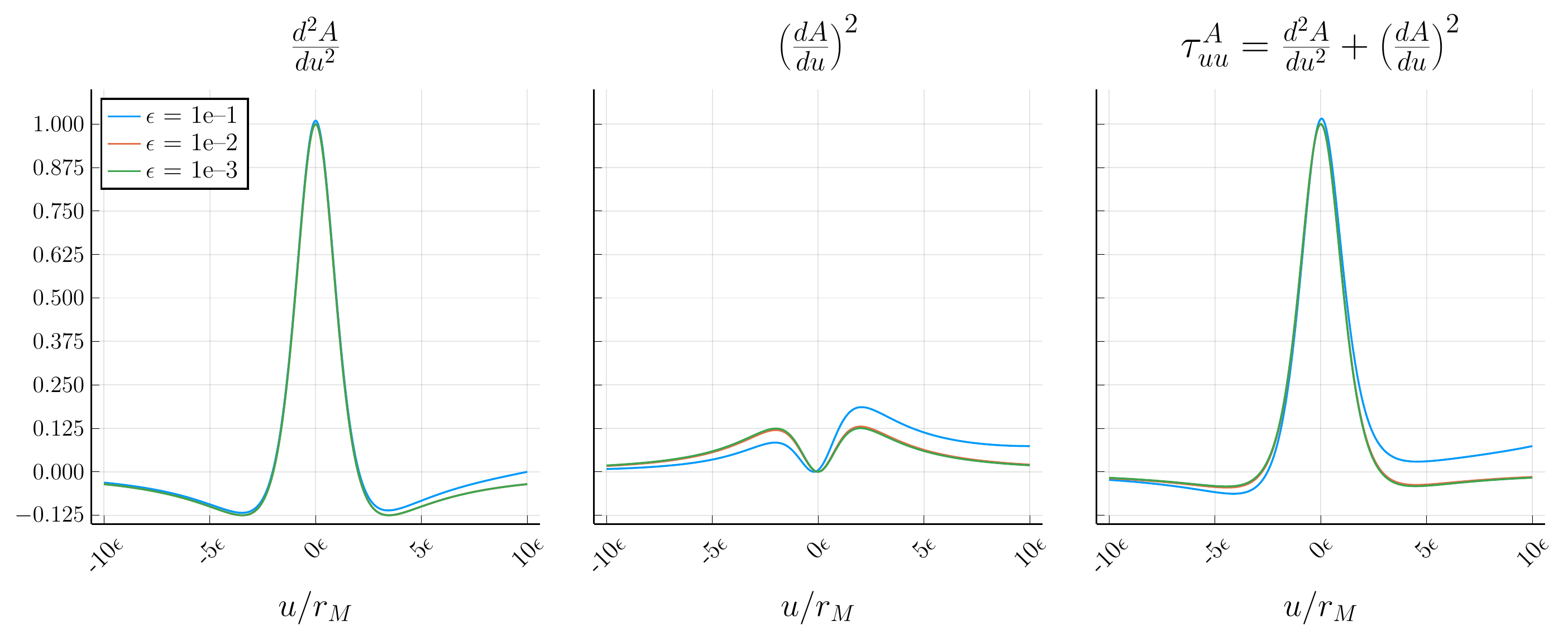}
\caption{First Two Panels: $\e^2 A''$ and $\e^2 A^{\prime\,2}$ of the regularized perturbation (\ref{Aeu}) 
as functions of $u$ in units of $1/8 \rM^2$ for $q=0$.  The horizon is at $u=0$, the $u-$axis is rescaled by $\e$ 
and the magnitude is rescaled by $\e^2$, showing that the self-similar behavior of the rescaled curves coincide 
for $\e \to 0$. Third Panel: The sum which contributes to (\ref{tauA}) and $\t_{uu}$ in units of $N\hbar/96\p\rM^2$,
also for $q=0$ and with axes similarly rescaled.}
\label{Fig:Tuu} 
\end{center}
\vspace{-1cm}
\end{figure}

The function $A'' + (A')^2$ which appears in  the quantum stress tensor (\ref{tauA}) has a maximum at 
$f_0 \sim \e \ll 1$ or at $u - (\vv_0-2\rM) \sim 2\,\e\, \rM$ with that maximum value there
of order $\e^{-2}$. The width in $u$ of the peak maximum in $A_{\e}$ is $\D u \sim 4 \rM \e$. 
The functions $A'', (A')^2$ and $A'' + (A')^2$ are plotted in Figs.~\ref{Fig:Tuu}. The main contribution 
comes from the region of $\D u\sim \e \rM$ around the maximum.
\vspace{3mm}

Since $f_0$ is a function of $u$, this effect is concentrated in an interval of $u$ near the horizon of order
\vspace{-3mm}
\be
\D u \sim \D r \sim \e \rM \sim \sqrt{N} L_{\rm Pl} 
\label{Delu}
\vspace{-3mm}
\ee
which is of the order or somewhat larger than the Planck scale $L_{\rm Pl}\equiv \sqrt{\hbar G/c^3} = 1.616 \times 10^{-33}$ cm., 
if we take $\e \sim \sqrt{N} L_{\rm Pl}/\rM$, which we shall show presently is the size needed for the quantum effects to significantly 
alter the classical geometry. Since $h\! =\! f(r) \!\to \!0$ for the Schwarzschild line element (\ref{Schw}), this corresponds to a {\it physical} 
distance scale of
\vspace{-3mm}
\be
\ell \sim \frac{\,\D r}{\!\!\!\sqrt{\e}} \sim N^{\frac{1}{4}} \sqrt{\rM L_{\rm Pl}} \gg L_{\rm Pl}
\label{elldef}
\ee
from the horizon. For a solar mass BH, $\ell$ is of order $10^{-14}$ cm or greater. Although very small
by astrophysical standards, since $\ell \gg L_{\rm Pl}$ by some $19$ orders of magnitude, one may still 
expect to be able to apply semi-classical methods in this regime. 

\begin{figure}[h]
\includegraphics[height=10cm,width=18cm, trim=3mm 0cm -5mm 0cm, clip=true]{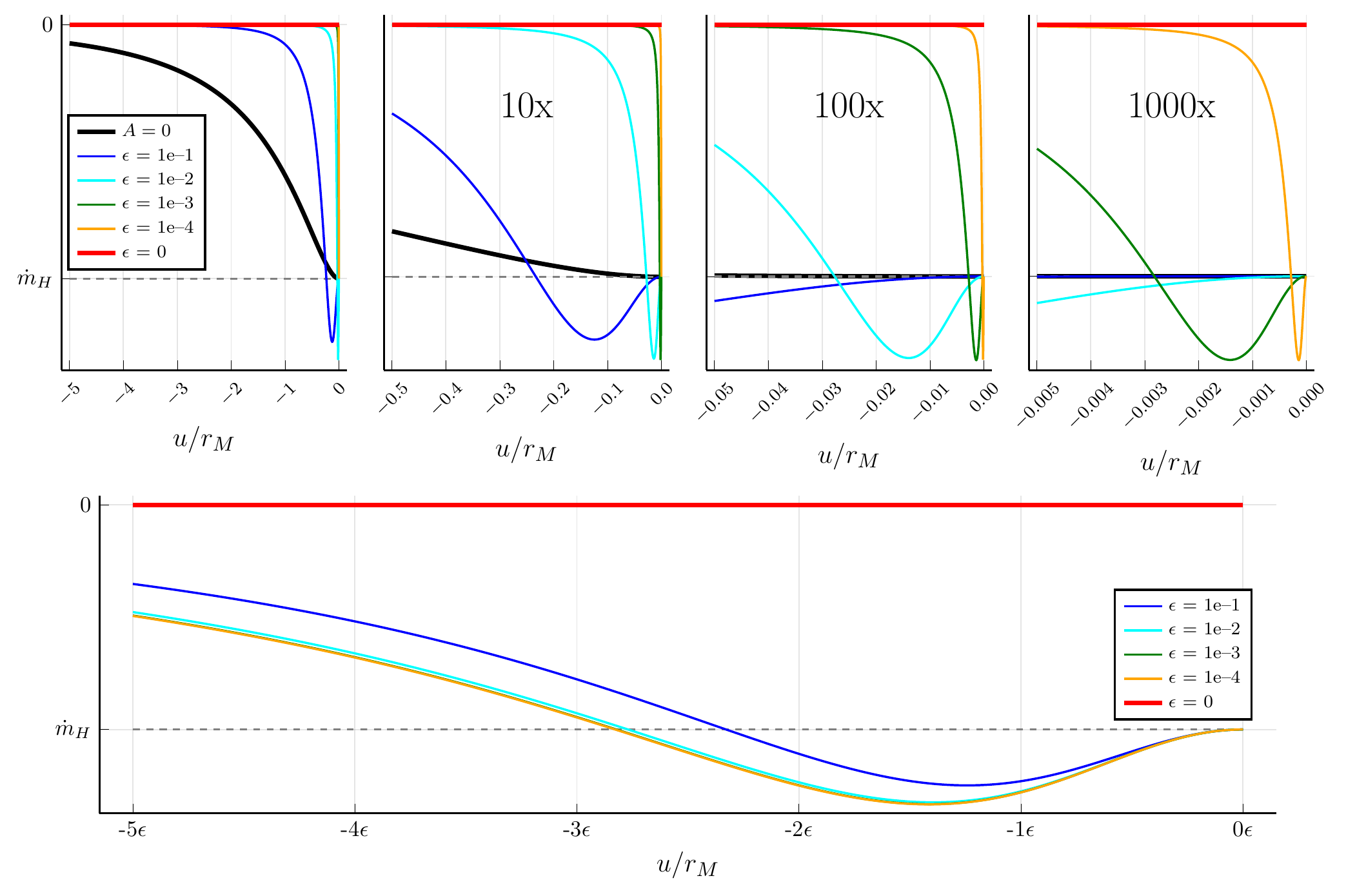}
\vspace{-8mm}
\caption{Upper Panels: Mass flux (\ref{dmdt}) as a function of horizon advanced time $u$, showing the suppression of the
Hawking flux by the perturbation $A_\e(u)$ in the initial state for $u$ increasingly close to the horizon at
$u=0$ for decreasing values of $\e$. $\dot m_H$ denotes the value of the 2D Hawking flux (\ref{Haw2D})
to which all regular perturbations tend finally at $u=0$.
Lower Panel: Expanded $u$ scale showing the self-similar behavior under rescalings of $\e$.} 
\label{Fig:Flux}
\vspace{-3mm}
\end{figure}

The behavior of the Hawking flux suppression for some moderately small values of $\e$ is illustrated in Fig.~\ref{Fig:Flux},
showing that this suppression persists for longer and longer retarded $u$ times closer to $u=\vv_0 -2\rM$ on the future horizon, 
for smaller and smaller $\e$. Given (\ref{regionII}) and (\ref{uutilde}), this corresponds at fixed $r$ to times $t \propto \rM \ln (1/\e)$
after the collapse of the null shell. Fig.~\ref{Fig:Flux} also exhibits the self-similar behavior of the flux suppression
as $u\!\to\! \vv_0-2\rM$ for $\e\!\to\! 0$, which is a consequence of the conformal properties of the spacetime in
near-horizon region~\cite{SacSol:2001,MazEMWeyl:2001,AntMazEM:2012}.

For a quantitative estimate of how large the effects of the perturbation (\ref{Aeu}) on the geometry would be,
if backreaction were to be taken into account, note that the overall scale of the quantum effects encoded in 
$\t_{ab}$ are of order $N\hbar/48 \p\rM^2$. From the four-dimensional Einstein tensor (\ref{Ein4}) and
stress tensor (\ref{t2D}), $\t_{ab}$ leads to effects on $G_{ab}$ of order $(8\pi G/4 \p r^2)\,\t_{ab}$, or
$N\hbar G/24 \p \rM^4$. This is to be compared with the 4D classical curvature components computed 
in the Schwarzschild geometry, given in Appendix \ref{App:Curv} which are of order $1/\rM^2$ at the horizon. 
Thus the quantum backreaction effects are generally suppressed by an overall relative factor of
\vspace{-3mm}
\be
\a_G \equiv \frac{N \hbar G}{24 \p \rM^2} = \frac{N\!}{24 \p}\,\left( \frac{L_{Pl}}{\rM} \right)^2 \ll 1
\label{supp}
\vspace{-1mm}
\ee
compared to the classical geometry.
This is certainly a very substantial suppression for a macroscopically large BH compared to the Planck scale, and the reason 
that quantum effects in classical GR are generally considered to be quite negligible. However even such an enormous 
suppression factor as (\ref{supp}) can be overcome if the quantum stress-tensor (\ref{Tuv}) components become 
large enough (while still remaining finite) in the vicinity of the future apparent horizon.

With (\ref{Aeu}) as a complete regularization of the non-vacuum initial state perturbation (\ref{Au}) in both regions,
the $A'' + (A')^2$ term in (\ref{tauA}) is of order $\e^{-2}$ in the near horizon region and the quantum suppression (\ref{supp})
is overcome if
\vspace{-3mm}
\be
\frac{\a_G}{\e^2} \,\left(q^2 -\sdfrac{1}{4}\right)  \ \gtrsim 1\,,\qquad {\rm or}\qquad \e \
\lesssim \ \text{Max}(1, |q|)\times  \sqrt{\frac{N}{6 \p}}\, \left(\frac{L_{\rm Pl}}{2 \rM}\right)\,.
\label{epssize}
\vspace{-1mm}
\ee
For large $|q| \gg 1$ the condition on how small $\e$ must be to overcome the suppression of quantum non-vacuum 
effects on the horizon is weakened by the appearance of a large factor of $|q|$ in (\ref{epssize}), but in
the following we assume that $q$ is of order $1$ and not particularly large, which we show in Sec.~\ref{Sec:Prob}
has the highest probability of occuring in the initial state.

Since the finite regularized perturbation (\ref{Aeu}) is present in the initial state, prior to the formation 
of the BH so we also estimate its total Misner-Sharp energy in the flat space region I where $R=0$ and 
(\ref{regionI}) applies. Using (\ref{dmdudv2}) and (\ref{Tuv}) with (\ref{Delu}) gives
\be
m = \int_{-\infty}^{\infty}  du\,  \frac{\pa m}{\pa u} =  \int_{-\infty}^{\infty}  du\,\t_{uu}
\sim \D u\,  \frac{\hbar N}{24 \p (\D u)^2} \sim  \frac{\hbar N}{24 \p \e \rM} 
\sim \!\!\sqrt{\sdfrac{N}{6 \p}}\,\sdfrac {M_{\rm Pl}}{2} \ll M
\label{mquan}
\ee
of the order the Planck mass $M_{\rm Pl} = 2.177 \times 10^{-5}$ gm. In the flat region $\D u \sim L_{\rm Pl}$, so that a quantum perturbation 
on the future apparent horizon of the BH large enough to overcome the suppression (\ref{supp}) and produce significant 
backreaction on the classical geometry only requires a Planck mass-energy fluctuation $M_{\rm Pl}$ concentrated within 
a Planck length $L_{\rm Pl}$ distance, just the scale at which such quantum fluctuations in the initial state are expected 
on general grounds of the uncertainty principle. 

In the next section we give a quantitative estimate of  the probability that such a non-vacuum quantum fluctuation 
large enough to satisfy the conditions (\ref{epssize})-(\ref{mquan}) exists in the wave functional of the initial vacuum state.

\section{Probability Distribution for Non-Vacuum Initial Conditions}
\label{Sec:Prob}

The effective action of the conformal anomaly (\ref{act2}) is quadratic in the conformalon scalar field $\vf$,
and its eq.~of motion (\ref{eom2}) in the asymptotically flat region where $R=0$ is that of a free scalar field. 
Since in a free theory the wave functional of the ground state vacuum is a simple Gaussian, evaluating the
width of this Gaussian enables us to give a quantitative estimate of the probability of the coherent state perturbation 
of the form of (\ref{Aeu}) parametrized by $\e$ and $q$.

For one simple harmonic oscillator with frequency $\w$, the classical action
\be
S_{\!\rm osc}[x] = \sdfrac{1}{2} \int dt\,  \left( \dot x^2 - \w^2 x^2\right) 
\ee
is quadratic in $x$, and the ground state of the oscillator is described by the Schr\"odinger wave function 
\be
\j_0(x) = \left(\frac{\w}{\p \hbar}\right)^{\!\frac{1}{4}} \exp \left(- \frac{\w x^2\!}{\!2\hbar}\,\right)
\label{hoscpsi}
\ee
which is a simple Gaussian, normalized to $\int_{-\infty}^{\infty} dx \,|\j_0(x)|^2 = 1$. Since $dx \,|\j_0(x)|^2$ is the 
probability of finding the oscillator with a value of the coordinate between $x$ and $x + dx$,  the probability of 
finding the coordinate $x$ with any absolute value $|x| \ge \bar x > 0$ is
\be
P (\bar x) = 2 \!\int_{\bar x}^\infty  dx\, |\j_0(x)|^2 = \text{erfc} \left(\!\!\sqrt{\frac{\w}{\hbar}}\, \bar x\right)
\label{Pxbar}
\ee
in terms of the complementary error function erfc.

This simple result can be generalized to a free QFT, viewed as a collection of free harmonic oscillators, in both the fixed 
time and light cone quantization schemes.  For initial data on a lightlike null surface such as $\sI^-$ the Schr\"odinger 
wave functional formulation is given in \cite{Heinzl:2001}. The Gaussian wave functional on the initial data for a 
canonically normalized scalar field $\f$ is proportional to
\be
\exp\left\{ -\sdfrac{1}{\hbar}\,\big(\f^-,\W \f^+\big)\right\}
\label{freewv}
\ee
where $\f^{\pm}$ are the positive and negative frequency parts of $\f$, and $\W = 2 k$, the analog of $\w$ 
in (\ref{hoscpsi}), is called the `covariance' and given in momentum space with $k$ the momentum conjugate
to the light front variable $u$ or $\vv$. For a real scalar field the positive and negative frequency parts are 
simply related by complex conjugation, {\it i.e.} $\f^- = (\f^+)^*$. Applying this general result to the 
anomaly effective action (\ref{act2}), the square of the ground state Schr\"odinger wave functional for the conformalon 
scalar $\vf$ on an initial null hypersurface is
\be
\big\vert\Psi_0[\vf]\big\vert^2 \propto \exp\left\{ - \frac{N}{24\p} \int_0^\infty \frac{dk}{2 \p}\  \vf^-(k)\, (2k)\, \vf^+(k)\right\}
\label{Gaussnull}
\ee
after account is taken of the normalization of (\ref{act2}) with the factor of $N\hbar/48\p$ relative to the canonical 
normalization of $1/2$ for a free scalar field. The overall normalization factor in (\ref{Gaussnull}) is to be determined by the
requirement that $|\Psi_0|^2$ integrated over all values of the parameters characterizing the initial state 
perturbation is $\vf$ is unity.

For the unregularized perturbation $\vf = 2 A(u)$ with $A(u)$ given by (\ref{Au}), the positive frequency component 
in momentum space is
\vspace{-1mm}
\be
\vf^+(k) = (2q +1) \int_{-\infty}^{\infty} \! du\ e^{iku} \ln |f_0|\,,\qquad k> 0\,.
\label{posphi}
\ee
which is the result of the $\e \to 0$ limit of the regularized form (\ref{Aeu}). With  the change of variables
$u=\vv_0 - 2\rM x$, (\ref{posphi}) is
\vspace{-3mm}
\be
\vf^+(k)  = 2\rM\,(2q+1)\, e^{ik\vv_0}\, I(z)\big\vert_{z=2k\rM}
\label{vfk}
\ee
where the integral $I(z)$ is finite and given by
\begin{align}
I(z) = \!\int_{-\infty}^\infty\!\!  dx \,e^{-ixz}\,  \ln\left\vert 1 - \sdfrac{1}{x}\right\vert
&= \int_1^\infty\!\!  dx \,e^{-ixz}\,  \ln\left( 1 - \sdfrac{1}{x}\right) +
\int_0^1 \!\!  dx \,e^{-ixz}\,  \ln\left( \sdfrac{1}{x}-1 \right) 
+ \int_0^{\infty}\!\!  dx \,e^{ixz}\,  \ln\left( 1 +\sdfrac{1}{x}\right)\nn
&= \frac{\p}{z}\left(1 -e^{-iz}\right)\,.
\label{Iz}
\end{align}
Although each of the three integrals in (\ref{Iz}) involves sine-integral (Si) and cosine-integral (Ci)
special functions, their sum turns out to be expressible in terms of elementary functions in the last form.

Substituting (\ref{vfk}) with (\ref{Iz}) and $z=2k\rM$ into (\ref{Gaussnull}) gives the probability density 
of the initial state perturbation
\be
\big\vert\Psi_0\big\vert^2 \propto
\exp\left\{-\frac{N}{24 \p^2} \, \big(2q+1\big)^2\! \int_0^\infty\! dz\ z\  \vert I(z)\vert^2\right\}
\label{Prob}
\ee
for the unregularized initial state perturbation (\ref{Au}). Now observe from (\ref{Iz}) that the integrand 
of the $z$ integral in (\ref{Prob}) is
\vspace{-1mm}
\be
z\  \vert I(z) \vert^2 = z \ \sdfrac{\p^2}{z^2}\left\vert 1 -e^{-iz}\right\vert^2 =
 \sdfrac{4\p^2}{z}\, \sin^2\left(\sdfrac{z}{2}\right) \sim \sdfrac{2\p^2}{z}
\label{asymp}
\ee
so that in fact the integral in (\ref{Prob}) as it stands diverges logarithmically, and would give an 
identically zero probability for any $q \neq -1/2$, which is the vacuum state. This is consistent with 
the general theorem of Ref.~\cite{FulSweWald:CMP78}, which excludes the possibility that truly singular behavior 
on the future horizon could be generated in gravitational collapse, starting from smooth initial data.
The perturbation (\ref{Au}) is such a singular perturbation for any $q\neq -1/2$, also with diverging 
energy (\ref{mquan}) in the initial state.

It is not difficult to see that the large $z$ behavior of the integral (\ref{Iz}) is determined by the behavior of $A(u)$
at its logarithmic singular points where $f_0$ becomes either $0$ or $\infty$. Thus if we replace the singular 
perturbation (\ref{Au}) by the finite one (\ref{Aeu}) regularized by a small but finite $\e$ parameter,
the $1/z$ behavior of (\ref{Iz}) and (\ref{asymp}) is cut off at $z \sim 1/\e$ with the result that
\vspace{-1mm}
\be
\int_0^\infty dz \, z\, |I_\e(z)|^2 \sim \ln \big(1/\e\big)
\vspace{-1mm}
\ee
for the regularized perturbation (\ref{Aeu}), and as a result the probability functional (\ref{Prob}) becomes
\vspace{-2mm}
\be
\vert\j_\e(q)\vert^2 \propto  \exp\left\{ - \frac{N}{24\p^2}\, (2q +1)^2 \,\ln \big(1/\e\big)\right\} =
\e^{N (2 q +1)^2/24\p^2}
\label{psiq}
\vspace{-1mm}
\ee
which is now a finite normalizable probability density in $q$ and for any $\e >0$. 

If $\e$ is required to satisfy (\ref{epssize}) for $q$ of order unity, it is instructive to evaluate the exponent for 
a typical value of $\rM \simeq 3$ km for a solar mass BH, for which
\vspace{-1mm}
\be
\frac{\rM}{L_{\rm Pl}} \simeq 1.9 \times 10^{38} \gg 1\,. 
\label{ratio}
\vspace{-1mm}
\ee
Despite this very large value, the exponent in (\ref{psiq}) is only weakly logarithimically dependent on $\e$ and
\vspace{-1mm}
\be
\frac{1}{24\p^2} \,\ln \big(1/\e\big) = 
\frac{1}{24\p^2} \,\ln \left(\!\!\sqrt{\frac{6\p}{N}}\ \frac{2\rM}{L_{\rm Pl}}\right) \simeq 0.38 - \frac{\ln N}{48\p^2}
\label{expln}
\ee
is actually $\cO(1)$. The $\ln N$ term in (\ref{expln}) is also negligibly small compared to $0.38$ provided
$\ln N \ll (48\p^2) (0.38) \simeq 180 $, so that neglecting it, we find from (\ref{psiq}) the normalized 
probability distribution in $q$ is approximately
\be
\vert\j_\e(q)\vert^2 \simeq \sqrt{\sdfrac{(1.52)\,N}{\p}}\, \exp\left\{ - (0.38) \, N \, (2q +1)^2\right\}
\label{psinorm}
\ee
centered on the vacuum value of $q=-\frac{1}{2}$, where the normalization is now fixed by 
$\int_{-\infty}^{\infty}\!dq\,\vert\j_\e(q)\vert^2 =1$.

For the perturbation (\ref{Aeu}) with $q=0$ that produces a large suppression of the Hawking effect and stress tensor 
on the horizon that is also large enough to produce significant backreaction according to (\ref{epssize}), we have
\be
\vert\j_\e(0)\vert^2 \simeq (0.70) \sqrt{N}\, e^{- (0.38) \, N} = (0.70) \sqrt{N}\, (0.68)^N
\label{Pest}
\ee
which is $\cO(1)$, unless $N$ is very large. As in (\ref{Pxbar}), the probability of finding a perturbation 
in the initial state of the form (\ref{Au}) varying from the vacuum value by $|\D q|\ge 1/2$ is
\be
P\left(|\D q| \ge\sdfrac{1}{2}\right) \simeq {\rm erfc}\left(\!\!\sqrt{(0.38)\, N}\right) 
= \left\{\begin{array}{cc} 0.38\,, & N=1\\ 0.08\,, &N=4\end{array}\right.
\label{P0N}
\ee
which is also $\cO(1)$, for $N$ fields contributing to the 2D conformal anomaly, unless $N$ is very large.

\section{Discussion and Outlook}
\label{Sec:Concl}

In this paper we have considered a simple 2D model of gravitational collapse, and studied the effects of 
the quantum conformal anomaly on the resulting classical BH horizon. Although this and similar 2D models
of gravitational collapse have been considered previously \cite{Unruh:1976,DavFulUnrPRD76,ParPirPRL94},
attention has been focused almost exclusively on initial conditions corresponding to the Minkowski vacuum 
on $\sI^-$. This choice of initial state leads to the stress tensor on the horizon that is regular in free-falling
coordinates and backreaction effects of Hawking radiation that are small, at least initially in the semi-classical 
approximation, where quantum fluctuations from the mean $\lag T_{\!\m}^{\ \n}\rag$ are ignored.

This study shows instead that the quantum effects of the conformal anomaly can be extraordinarily large on BH horizons, 
overcoming even the enormous suppression of Planck to macroscopic scales expressed by the ratio (\ref{ratio}). This 
suppression, normally expected of quantum effects in classical gravity, can be overcome in the stress tensor of the conformal
anomaly because of its sensitivity to light cone pole singularities of quantum field theory, that occur in generic quantum states 
and extend to macroscopic scales. This specifically quantum, non-local effect, and its importance to the behavior of the stress 
tensor on BH horizons is illustrated in the simple 2D model of this paper.

This is a proof of principle of state-dependent anomaly effects on BH horizons in a simple 2D model with $p_\perp\!=\!0$.
Relaxing this condition to obtain a more realistic model will require use of the full 4D conformal anomaly effective action and 
stress tensor of~\cite{EMVau:2006,EMEFT:2022}, which nonetheless is expected to have similar significant state-dependent 
effects on the future event horizon of a 4D black hole, as already pointed out in~\cite{EMVau:2006}. The present paper 
therefore provides a good motivation and warm-up for study of the more realistic but technically more challenging 4D collapse 
problem by similar methods applied to the 4D anomaly stress tensor. The significant effects of the conformal anomaly
even in the simplified 2D model of this paper support the conclusion that the effective action of the conformal anomaly
is a relevant addition to the classical theory that should be added in a full effective field theory (EFT) treatment of gravity 
at macroscopic scales~\cite{EMVau:2006,GiaEM:2009,EMZak:2010,EMSGW:2017,EMEFT:2022}.

By recasting the effective action of the 2D conformal anomaly in local form (\ref{act2}) via the introduction 
of a local scalar conformalon field $\vf$, a very wide class of initial conditions can be considered, by 
allowing general homogeneous solutions to the linear wave eq.~(\ref{eom2}) that $\vf$ satisfies. As a 
practical matter, this formulation of general initial conditions is simpler and much less technically
involved than calculating the stress tensor of every quantum field in each and every quantum state,
by the standard approach of mode sums, which requires a cumbersome process of regularization and 
renormalization on a case by case basis, even on a fixed background with a great deal of symmetry~\cite{BirDav}. 
Calculations of quantum backreaction in dynamically evolving spacetimes, or those with less symmetry
rapidly become prohibitive by this method. The local form of the conformal anomaly stress tensor
and eq.~of motion provides a more practical approach to make progress in this class of 
quantum backreaction problems in BH and other curved spacetimes, particularly in the horizon region
where the anomaly dominates other vacuum polarization effects because of its lightlike singularity. 

The relevance of the anomaly stress tensor in the 2D model is illustrated through its effect on Hawking emission, 
which can be modified or suppressed for indefinitely long times after gravitational collapse, by different choices 
of the initial state easily studied by means of different homogeneous solutions to the $\vf$ eq.~(\ref{eom2}).
Since the anomaly effective action is quadratic in $\vf$, it is also a convenient route to estimating the
probability of such non-vacuum initial conditions in the vacuum wave functional. The probability
of non-vacuum initial conditions that can significantly affect the BH near-horizon geometry and Hawking
effect (\ref{Pest})-(\ref{P0N}) are not negligibly small, but rather of $\cO(1)$. This demonstrates
the ability of the anomaly to overcome large quantum suppression factors in gravitational 
collapse, and the special and fine-tuned nature of the vacuum initial conditions upon which virtually all inferences 
of quantum effects in BHs have been based. The present study indicates that a reconsideration of these
conclusions for more general initial state conditions is warranted.

Clearly the estimates of the probability based on a 2D model of gravitational collapse (\ref{Pest})-(\ref{P0N}) are only 
illustrative, given that the 2D model itself is incomplete, by setting to zero identically the transverse pressure
as in (\ref{Ein2pb}). The shortcomings of this model and similar ones have been pointed out~\cite{BalbFabPRD99}. For these 
reasons we do not take (\ref{Pest})-(\ref{P0N}) as accurate reliable predictions for the probability of non-vacuum initial conditions 
in 4D gravitational collapse. Nevertheless, general features of weak, logarithmic dependence on the large ratio of scales
$1/\e \sim \rM/L_{\rm Pl}$ of this probability function, when initial state perturbations are regularized by a small 
parameter that grow large on the horizon, are expected to hold in four dimensions as well. The 4D effective action
of the conformal anomaly is also quadratic in $\vf$, and its eq.~of motion is also linear~\cite{EMSGW:2017,EMEFT:2022}.
Hence the probability of non-vacuum initial conditions that lead to large effects on the BH horizon found
in~\cite{EMVau:2006,EMZak:2010} can be studied by the same methods as those in the 2D case.
Thus the study of the simplified 2D model presented here justifies a detailed study of the analogous 
non-vacuum perturbations by means of the 4D quantum conformal anomaly in more realistic models of gravitational 
collapse, and in the full EFT of \cite{EMEFT:2022}, where the $\vf$ conformalon is coupled to dynamical vacuum energy, 
allowing it also to change in the near-horizon region, and possibly leading to a regular de Sitter interior consistent
with quantum theory~\cite{EMSpringer:2023}.

\pagebreak

\bibliographystyle{apsrev4-1}
\bibliography{gravity21Aug}

\appendix

\section{Curvature Components in Double Null Coordinates}
\label{App:Curv}

\numberwithin{equation}{section}

\setcounter{equation}{0}

To calculate the Riemann curvature components most rapidly we use the method of differential forms and the
definition of the vierbeine or tetrad frame one-forms
\vspace{-3mm}
\be
e^{\ha} = e^{\ha}_{\ \m}\, dx^\m
\vspace{-2mm}
\ee
in the orthornormal coordinates denoted by the hatted indices, such that the metric can be written
\vspace{-3mm}
\be
g_{\m\n} = \h_{\ha\hb}\, e^{\ha}_{\ \m} e^{\hb}_{\ \n}
\vspace{-2mm}
\ee
in terms of the constant metric and its inverse
\vspace{-3mm}
\be
\h_{\hu\hv} = \h_{\hv\hu} =-\sdfrac{1}{2}\,,\qquad \h^{\hu\hv} = \h^{\hv\hu} =-2\,,\qquad
\h_{\hth\hth} = \h_{\hf\hf}= \h^{\hth\hth} = \h^{\hf\hf} = 1\,.
\label{hatgflat}
\vspace{-1mm}
\ee
In flat space $r = (\vv-u)/2$, whereas in the general spherical symmetric geometry in double null coordinates
the metric is given (\ref{gensphsym}) and (\ref{doubnull}), in terms of two functions $r(u,\vv)$ and
$\s(u,\vv)$ to be determined. Therefore we may choose the frame one-forms and vierbein fields to be
\vspace{-2mm}
\bes
\bea
&e^{\hu} = e^\s\, du\,,\qquad e^{\hu}_{\ u}= e^\s\\
&e^{\hv} = e^\s\, d\vv\,,\qquad e^{\hv}_{\ \vv}= e^\s\\
&e^{\hth} = r\, d\th\,,\qquad e^{\hth}_{\ \th}= r\\
&e^{\hf} = r \sin\th\, d\f\,,\qquad e^{\hf}_{\ \f}= r\sin\th
\eea
\label{tetrads}\ees
with all other components $e^{\ha}_{\ \m}$ not listed in the second column vanishing.

From the above frame one-forms the connection one-forms $w^{\ha}_{\ \hb}$ are determined by
the requirement from Cartan's second eq.~of structure
\vspace{-3mm}
\be
\cT^{\ha} \equiv d e^{\ha} + w^{\ha}_{\ \hb} \wedge e^{\hb}= 0
\label{zerotor}
\vspace{-1mm}
\ee
of vanishing torsion $\cT^{\ha}$. Here $d$ here denotes exterior differentiation of forms and the $\wedge$ (`wedge') operation
denotes the anti-symmetric product of forms. Thus from (\ref{tetrads}), (\ref{zerotor}) and 
\vspace{-2mm}
\bes
\bea
&de^{\hu} = -e^\s\, \pa_\vv \s\,  du \wedge d\vv\\
&de^{\hv} = +e^\s\, \pa_u \s\, du \wedge d\vv\\
&de^{\hth} = +\pa_u r \,du \wedge d\th +\pa_\vv r\, d\vv \wedge d\th \\
&de^{\hf} = +\sin\th\, \pa_u r \, du \wedge d\f+\sin\th\, \pa_\vv r \, d\vv \wedge d\f  + r\cos\th \,d\th\wedge d\f
\eea
\label{dtetrad}\ees
\vspace{-2mm}
one finds
\vspace{-4mm}
\bes
\bea
&w^{\hu}_{\ \hu} = -w^{\hv}_{\ \hv} = \pa_u \s\,  du - \pa_\vv \s\,  d\vv  \\
&w^{\hu}_{\ \hv}  = w^{\hv}_{\ \hu}  = 0\\
&w^{\hu}_{\ \hth} = 2w^{\hth}_{\ \hv} = 2e^{-\s} \, \pa_\vv r\, d\th   \\
&w^{\hv}_{\ \hth} = 2w^{\hth}_{\ \hu} = 2e^{-\s} \, \pa_u r\, d\th  \\
&w^{\hu}_{\ \hf} = 2 w^{\hf}_{\ \hv}  = 2e^{-\s}\sin\th  \, \pa_\vv r\, d\f \\
&w^{\hv}_{\ \hf} = 2w^{\hf}_{\ \hu} = 2e^{-\s}\sin\th \, \pa_u r\, d\f  \\
&w^{\hf}_{\ \hth} = - w^{\hth}_{\ \hf} = \cos\th \,d\f 
\eea
\ees
for the connection one-forms, with terms not listed vanishing.

The Riemann curvature two-form is then calculated from Cartan's first eq.~of structure
\vspace{-2mm}
\be
\cR^{\ha}_{\ \hb} \equiv d w^{\ha}_{\ \hb} + w^{\ha}_{\ \hc} \wedge w^{\hc}_{\ \hb} 
= R^{\ha}_{\ \hb\hc\hd} \,e^{\hc}\wedge e^{\hd}
\label{Rieform}
\ee
from which we obtain the $20$ non-vanishing components of the Riemann tensor
\vspace{-2mm} 
\bes
\bea
&R^{\hu}_{\ \hu\hv\hu}= R^{\hv}_{\ \hv\hu\hv}= 2\,e^{-2\s}\, \pa_{u}\pa_{\vv}\s\\
&R^{\hu}_{\ \hth \hu\hth} =R^{\hu}_{\ \hf \hu\hf}=R^{\hv}_{\ \hth \hv\hth}= R^{\hv}_{\ \hf \hv\hf}
=2 R^{\hth}_{\ \hu \hv\hth} =2 R^{\hth}_{\ \hv \hu\hth} =2 R^{\hf}_{\ \hu \hv\hf} =2 R^{\hf}_{\ \hv \hu\hf} 
=\displaystyle{\sdfrac{2}{r}}\,e^{-2\s}\, \pa_{u}\pa_{\vv} r\\
&R^{\hu}_{\ \hth \hv\hth} = R^{\hu}_{\ \hf \hv \hf}= 2 R^{\hth}_{\ \hv \hv \hth}= 2 R^{\hf}_{\ \hv \hv \hf}
=\displaystyle{\sdfrac{2}{r}}\,e^{-2\s}\, \left(\pa_{\vv}^2 r\ -2\, \pa_{\vv}r\,\pa_{\vv}\s\right)\\
&R^{\hv}_{\ \hth \hu\hth} =R^{\hv}_{\ \hf \hu\hf} =2 R^{\hth}_{\ \hu \hu\hth} =2 R^{\hf}_{\ \hu \hu\hf}
=\displaystyle{\sdfrac{2}{r}}\,e^{-2\s}\, \left(\pa_{u}^2 r\ -2\, \pa_{u} r\,\pa_{u}\s\right)\\
&R^{\hth}_{\ \hf\hth\hf} = R^{\hf}_{\ \hth\hf\hth} =
\displaystyle{\sdfrac{1}{r^2}}\left(1 + 4\, e^{-2\s} \, \pa_{u} r \, \pa_{\vv} r\right)
\eea
\ees
in the orthonormal basis, together with the $20$ components related to these by anti-symmetry in the last two indices:
$R^{\ha}_{\ \hb \hc\hd} =  -R^{\ha}_{\ \hb\hd\hc}$.

The non-vanishing components of the Ricci tensor are 
\bes
\bea
&R^{\hu}_{\ \hu}= R^{\hv}_{\ \hv}= R^{u}_{\ u}= R^{\vv}_{\ \vv}=4\,e^{-2\s}\,\left( \pa_{u}\pa_{\vv}\s + 
\displaystyle{\sdfrac{1}{r}}\, \pa_{u}\pa_{\vv} r\right)\\
&R^{\hu}_{\ \hv}= R^{u}_{\ \vv}=\displaystyle{\sdfrac{4}{r}}\,e^{-2\s}\, \left(\pa_{\vv}^2 r\ -2\, \pa_{\vv}r\,\pa_{\vv}\s\right)\\
&R^{\hv}_{\ \hu}= R^{\vv}_{\ u}=\displaystyle{\sdfrac{4}{r}}\,e^{-2\s}\, \left(\pa_{u}^2 r\ -2\, \pa_{u}r\,\pa_{u}\s\right)\\
&R^{\hth}_{\ \hth} = R^{\hf}_{\ \hf} = R^{\th}_{\ \th} = R^{\f}_{\ \f} = 
4\, e^{-2\s} \, \left(\displaystyle{\sdfrac{1}{r}}\,\pa_{u} \pa_{\vv} r
+ \displaystyle{\sdfrac{1}{r^2}}\,\pa_{u} r \, \pa_{\vv} r\right) +\displaystyle{\sdfrac{1}{r^2}}
\eea
\ees
given in both the orthonormal and coordinate bases. Thus the four-dimensional Ricci scalar is
\vspace{-2mm}
\be
^{(4)}R = 8\,e^{-2\s}\,\left( \pa_{u}\pa_{\vv}\s + 
\displaystyle{\sdfrac{2}{r}}\,\pa_{u} \pa_{\vv} r
+ \displaystyle{\sdfrac{1}{r^2}}\,\pa_{u} r \, \pa_{\vv} r\right) +\displaystyle{\sdfrac{2}{\,r^2}}
\ee
and the non-vanishing components of the Einstein tensor are
\vspace{-2mm}
\bes
\bea
&G^{u}_{\ u}= G^{\vv}_{\ \vv}= - 4\,e^{-2\s}\,\left(\displaystyle{\sdfrac{1}{r}}\,\pa_{u} \pa_{\vv} r
+ \displaystyle{\sdfrac{1}{r^2}}\,\pa_{u} r \, \pa_{\vv} r\right) -\displaystyle{\sdfrac{1}{r^2}}\\
&G^{u}_{\ \vv}=\displaystyle{\sdfrac{4}{r}}\,e^{-2\s}\, \left(\pa_{\vv}^2 r\ -2\, \pa_{\vv}r\,\pa_{\vv}\s\right)\\
&G^{\vv}_{\ u}=\displaystyle{\sdfrac{4}{r}}\,e^{-2\s}\, \left(\pa_{u}^2 r\ -2\, \pa_{u}r\,\pa_{u}\s\right)\\
&G^{\th}_{\ \th} = G^{\f}_{\ \f} = -4\,e^{-2\s}\,\left( \pa_{u}\pa_{\vv}\s + 
\displaystyle{\sdfrac{1}{r}}\, \pa_{u}\pa_{\vv} r\right)
\eea
\ees
in the $(u,\vv,\th,\f)$ coordinate basis. All curvature components vanish for $\s=0, r= (\vv-u)/2$ in flat space.
With these results the Einstein eqs.~in the full four-dimensional space (\ref{gensphsym}) take the form
\vspace{-2mm}
\bes
\bea
&&\frac{\pa^2 r}{\pa u^2} 
-2\, \frac{\pa r}{\pa u} \frac{\pa \s}{\pa u} = -\frac{G}{r}\,\t_{uu}\,,\\
&&\frac{\pa^2 r}{\pa \vv^2} 
-2\, \frac{\pa r}{\pa \vv} \frac{\pa \s}{\pa \vv} = -\frac{G}{r}\,\t_{\vv\vv}\,,\\
&&\hspace{-2mm}\frac{\pa^2 r}{\pa u\pa v} 
+ \frac{1}{r} \frac{\pa r}{\pa u} \frac{\pa r}{\pa v} + \frac{e^{2\s}}{4r} = 
\frac{G}{r}\,\t_{uv}\,,\label{uveq}\\
&&\hspace{-5mm} \frac{\pa^2 \s}{\pa u\pa v} 
+ \frac{1}{r}\frac{\pa^2 r}{\pa u\pa v} = \sq \s + \frac{1}{r}\, \sq r = 0\,.
\label{Rcond}\eea
\label{Eindn}\ees

\vspace{-1cm}
\section{The Functions $r(u,\vv)$ and $\s(u,\vv)$ in regions I and II}
\label{App:Deriv}

In the flat region I, $\s=0$ and the $r= (\vv-u)/2$. Thus we have simply
\vspace{-1mm}
\bes
\begin{align}
\displaystyle{\frac{\pa r}{\pa \vv}}&=\displaystyle{ \frac{1}{2} = -\frac{\pa r}{\pa u}}\\
\displaystyle{\frac{\pa^2 r}{\pa \vv^2}=\frac{\pa^2 r}{\pa u^2}}&= \displaystyle{\frac{\pa^2 r}{\pa u\pa \vv} = 0}
\qquad {\rm in\ region\ I}.
\end{align}
\ees
In region II differentiation of (\ref{rstaruv}) gives
\vspace{-2mm}
\be
dr^* = \frac{dr}{f} = \frac{d\vv}{2} - \frac{d\tu}{2} = \frac{d\vv}{2} - \frac{du}{2f_0}
\ee
so that 
\vspace{-5mm}
\bes
\begin{align}
&\displaystyle{\frac{\pa r}{\pa u} = -\frac{f}{2f_0}}\\
&\displaystyle{\frac{\pa r}{\pa \vv} = \frac{f}{2}}\\
&\displaystyle{\frac{\pa^2 r}{\pa u^2} = \frac{ff'}{4f_0^2} - \frac{ff_0'}{4f_0^2}}\\
&\displaystyle{\frac{\pa^2 r}{\pa \vv^2}= \frac{ff'\!}{\!4}}\\
&\displaystyle{\frac{\pa^2 r}{\pa u\pa \vv}= - \frac{ff'\!}{4f_0}}\qquad \qquad{\rm in\ region\ II}
\end{align}
\ees
where
\vspace{-3mm} 
\be 
f' \equiv \frac{df}{dr} = \frac{\rM}{r^2}\,,\qquad f_0' \equiv f'\big\vert_{r=r_0} = \frac{\rM}{r_0^2}
\ee
so that $r, \pa_u r$ and $\pa_u^2 r$ are continuous at $\vv=\vv_0$, whereas
the $\vv$ derivatives and mixed $u,\vv$ second derivative of $r$ are not. 

From (\ref{sigmatch}), we also have in region II that
\vspace{-2mm}
\bes
\begin{align}
&\displaystyle{\frac{\pa \s}{\pa u} =- \frac{f'}{4f_0} + \frac{f'_0}{4 f_0}}\\
&\displaystyle{\frac{\pa \s}{\pa \vv} = \frac{f'\!}{\!4}}\\
&\displaystyle{\frac{\pa^2 \s}{\pa u\,\pa \vv} = -\frac{ff''\!}{\!8f_0}}\\
&\displaystyle{\frac{\pa^2 \s}{\pa u^2} = \frac{1}{8f_0^2} \left(f'' f - f_0f''_0 + f_0^{\prime\,2} - f' f_0'\right)}\\
&\displaystyle{\frac{\pa^2 \s}{\pa \vv^2} = \frac{ff''\!\!}{\!8}}\qquad \qquad{\rm in\ region\ II}
\end{align}\label{derivsig}
\ees
so that $\s$ and $\pa_u \s$ are continuous at $\vv=\vv_0$, whereas $\pa_\vv\s$ and $\pa_u\pa_\vv \s$ are not.

From these expressions one finds
\vspace{-3mm}
\be
G_{uu} = G_{u\vv} = G_{\th\th} = G_{\f\f} =0
\vspace{-2mm}
\ee
everywhere in both regions I and II, satisfying the vacuum Einstein eqs.

$G_{\vv\vv}$ also vanishes in each region I and II separately, but since
\be
\frac{\pa r}{\pa \vv} = \frac{1}{2}\, \Theta(\vv_0 -\vv) + \frac{f}{2}\, \Theta(\vv-\vv_0)
\ee
is discontinuous at $\vv=\vv_0$, its derivative
\be
\frac{\pa^2 r}{\pa \vv^2} = \frac{f-1}{2}\, \d (\vv-\vv_0) + \frac{f' f}{4} \Theta(\vv-\vv_0)
\ee
has a Dirac $\d$-function contribution, and
\be
G_{\vv\vv} = -\frac{2}{r} \left(\frac{\pa^2 r}{\pa \vv^2} -2\,  \frac{\pa r}{\pa \vv}\frac{\pa \s}{\pa \vv}\right)
= \frac{\rM}{r^2}\, \d(\vv-\vv_0) = \frac{2G \t_{\vv\vv}^{(C)}}{r^2}= 8 \p G T_{\vv\vv} 
\label{TGvvB}
\ee
evaluated at $\vv=\vv_0, r=r_0, f\!=\!f_0 \!\equiv\! f(r_0)$. Hence eq.~(\ref{TGvvB}), which is the only non-trivial Einstein 
eq.~due to the null shell is also satisfied and is (\ref{TGvv}) of the text.

Additionally, for the quantum anomaly stress tensor the required terms are
\bes
\begin{align}
\displaystyle{\frac{\pa^2 \s}{\pa u\,\pa \vv}} = \frac{\rM}{\!4r^3}\frac{f}{f_0}&\\
\displaystyle{\frac{\pa^2 \s}{\pa u^2} - \left(\frac{\pa \s}{\pa u}\right)^2} =
\frac{1}{16 f_0^2} \left(2ff'' -f^{\prime\,2} - 2f_0f_0'' + f_0^{\prime\,2}\right)
&= \frac{\rM}{4f_0^2} \left[ \frac{1}{r_0^3} - \frac{1}{r^3} 
+ \frac{3 \rM}{4} \left(\frac{1}{r^4} - \frac{1}{r_0^4}\right)\right]\\
\displaystyle{\frac{\pa^2 \s}{\pa \vv^2} - \left(\frac{\pa \s}{\pa \vv}\right)^2} =
\frac{1}{16} \left(2ff'' -f^{\prime\,2}\right) &= - \frac{\rM}{4r^3} \left(1 - \frac{3\rM\!}{\!4r}\,\right)
\end{align}
\ees
in the Schwarzshild region II.

\section{Three Sets of Double Null Coordinates and Horizon Finiteness Conditions}
\label{App:DNull}

We use two different sets of double null coordinates in this paper, which we designate $(u,\vv)$ and $(\tu, \tv)$.
A third set of Kruskal double null coordinates designated by $(U,V)$ are also often used for the
Schwarzschild solution. For the benefit of the reader we give here the relationships between the three different sets
of double null coordinates.

The first set are the simply double null coordinates in the flat region I before the passage of the null shell, defined
in (\ref{regionI}). The two other sets of coordinates are referred back and related to this first
and primary set of $(u,\vv)$ coordinates.

In crossing the imploding null shell at $\vv=\vv_0$ into region II we are in a Schwarzschild region with total mass
$M$ fixed by the null shell (\ref{Mdef}), (\ref{Edelta}). The Schwarzschild region II has metric and double null
Eddington-Finkelstein coordinates defined by (\ref{regionII}), and denoted $(\tu,\tv)$. In these Schwarzschild 
E-F coordinates one can find the solution to the $\vf$ eq. (\ref{eomtx})-(\ref{phisoln})
and see that it gives the diverging stress tensor stress tensor components (\ref{T2Dstat}). 

Since both sets of Schwarzschild $(t,r)$ and $(\tu,\tv)$ coordinates diverge at the horizon, one can
introduce Kruskal double null coordinates $(U,V)$ related to $(\tu,\tv)$ by
\bes
\begin{align}
U&= -2\rM \,e^{-\tu/2\rM}=-2\rM\, e^{-u/2\rM}\displaystyle{ \left(\sdfrac{r_0(u)}{\rM} -1 \right)}\\
V&= 2 \rM\, e^{\tv/2\rM}= 2 \rM\, e^{\vv/2\rM}\\
UV&= -4\rM^2\,e^{r^*/\rM} = -4 r\rM e^{r/\rM} f(r)
\end{align}
\label{Kruskal}\ees
which are regular on the horizon, mapping the future and past horizons to $U=0$ and $V=0$ respectively.
Thus the total Jacobian is
\be
\frac{dU}{du} = \frac{dU}{d\tu} \frac{d\tu}{du} = e^{-\tu/2\rM} \frac{1}{f_0} = e^{-u/2\rM} \frac{r_0(u)}{\rM}
\ee
showing that the total transformation from the original $(u,\vv)$ to Kruskal $(U,V)$ coordinates is non-singular
at $u = \vv_0 -2\rM, r_0(u) = \rM$ at the future classical horizon. Both these two sets of double null coordinates are
regular and horizon-penetrating on the future horizon, whereas the E-F $(\tu,\tv)$ are not.

The conditions of horizon regularity on the stress tensor are that all components are finite in any set
of coordinates that are non-singular on the horizon. Since both the Kruskal double null coordinates $(U,V)$
and flat double null coordinates $(u,\vv)$ of region I are non-singular on the horizon and
\vspace{-5mm}
\bes
\begin{align}
T_{\vv\vv} & = T_{\tv\tv}\\
T_{u\vv} & = \left(\frac{d\tu}{du}\right)\, T_{\tu\tv} =  \left(\sdfrac{1}{f_0}\right)\, T_{\tu\tv}\\
T_{uu} &= \left(\frac{d\tu}{du}\right)^2\,T_{\tu\tu} = \left(\sdfrac{1}{f_0}\right)^2\,T_{\tu\tu} 
\end{align}
\label{uvfinite}\ees
with (\ref{uutilde}), finiteness on the horizon requires each of the three components at left must be finite.
Since the ratio $f/f_0$ is finite on the horizon by (\ref{fratio}),  this implies
\vspace{-2mm}
\bes
\begin{align}
\lim_{r\to\rM} \, |T_{\tv\tv}| &<\infty\\
\lim_{r\to\rM} f^{-1} \,|T_{\tu\tv}| &< \infty\\
\lim_{r\to\rM} f^{-2} \,|T_{\tu\tu}| &< \infty
\end{align}
\ees
in agreement with Ref.~\cite{ChrFul:1977}.  These conditions are satisfied for the regularized initial state perturbation
(\ref{Aeu}) for $\e >0$.

\end{document}